\RequirePackage{fix-cm}
\documentclass[twocolumn]{svjour3}       
\smartqed  
\usepackage{graphicx}
\journalname{Computing and Software for Big Science}

\usepackage{ifthen} 
\newboolean{pdflatex}
\setboolean{pdflatex}{true} 

\newboolean{articletitles}
\setboolean{articletitles}{true} 

\newboolean{uprightparticles}
\setboolean{uprightparticles}{false} 

\newboolean{inbibliography}
\setboolean{inbibliography}{false} 

\usepackage{url}

\usepackage{subfig}

\usepackage{xspace} 
\usepackage{upgreek}

\def\MagUp {\mbox{\em Mag\kern -0.05em Up}\xspace}

\ifthenelse{\boolean{uprightparticles}}%
{

 \def\Pmu         {\ensuremath{\upmu}\xspace}

 \def\Ppi         {\ensuremath{\uppi}\xspace}

 \def\PDelta      {\ensuremath{\Delta}\xspace}                 
 \def\PXi      {\ensuremath{\Xi}\xspace}                 
 \def\PLambda      {\ensuremath{\Lambda}\xspace}                 
 \def\PSigma      {\ensuremath{\Sigma}\xspace}                 
 \def\POmega      {\ensuremath{\Omega}\xspace}                 
 \def\PUpsilon      {\ensuremath{\Upsilon}\xspace}                 
 
 \def\PB      {\ensuremath{\mathrm{B}}\xspace}                 
                  
 \def\PD      {\ensuremath{\mathrm{D}}\xspace}

 \def\PK      {\ensuremath{\mathrm{K}}\xspace}

 \def\Pi      {\ensuremath{\mathrm{i}}\xspace}

 \def\Ps      {\ensuremath{\mathrm{s}}\xspace}

}
{

 \def\Pmu         {\ensuremath{\mu}\xspace}

 \def\Ppi         {\ensuremath{\pi}\xspace}

 \mathchardef\PDelta="7101
 \mathchardef\PXi="7104
 \mathchardef\PLambda="7103
 \mathchardef\PSigma="7106
 \mathchardef\POmega="710A
 \mathchardef\PUpsilon="7107
                  
 \def\PB      {\ensuremath{B}\xspace}                 
                  
 \def\PD      {\ensuremath{D}\xspace}

 \def\PK      {\ensuremath{K}\xspace}

 \def\Pi      {\ensuremath{i}\xspace}

 \def\Ps      {\ensuremath{s}\xspace}

}

\makeatletter
\ifcase \@ptsize \relax
  \newcommand{\miniscule}{\@setfontsize\miniscule{4}{5}}
\or
  \newcommand{\miniscule}{\@setfontsize\miniscule{5}{6}}
\or
  \newcommand{\miniscule}{\@setfontsize\miniscule{5}{6}}
\fi
\makeatother

\DeclareRobustCommand{\optbar}[1]{\shortstack{{\miniscule (\rule[.5ex]{1.25em}{.18mm})}
  \\ [-.7ex] $#1$}}

\def\mumu       {{\ensuremath{\Pmu^+\Pmu^-}}\xspace}

\def\squark    {{\ensuremath{\Ps}}\xspace}

\def\pion   {{\ensuremath{\Ppi}}\xspace}

\def\pip    {{\ensuremath{\pion^+}}\xspace}

\def\kaon    {{\ensuremath{\PK}}\xspace}
  \def\Kbar    {{\kern 0.2em\overline{\kern -0.2em \PK}{}}\xspace}

\def\KorKbar    {\kern 0.18em\optbar{\kern -0.18em K}{}\xspace}

\def\Kp      {{\ensuremath{\kaon^+}}\xspace}
\def\Km      {{\ensuremath{\kaon^-}}\xspace}

\def\Kstarz  {{\ensuremath{\kaon^{*0}}}\xspace}

  \def\Dbar    {{\kern 0.2em\overline{\kern -0.2em \PD}{}}\xspace}

\def\DorDbar    {\kern 0.18em\optbar{\kern -0.18em D}{}\xspace}

\def\B       {{\ensuremath{\PB}}\xspace}
\def\Bbar    {{\ensuremath{\kern 0.18em\overline{\kern -0.18em \PB}{}}}\xspace}

\def\BorBbar    {\kern 0.18em\optbar{\kern -0.18em B}{}\xspace}
\def\Bz      {{\ensuremath{\B^0}}\xspace}

\def\Bs      {{\ensuremath{\B^0_\squark}}\xspace}

  \def\Y#1S{\ensuremath{\PUpsilon{(#1S)}}\xspace}

\def\Lbar        {{\ensuremath{\kern 0.1em\overline{\kern -0.1em\PLambda}}}\xspace}
\def\LorLbar    {\kern 0.18em\optbar{\kern -0.18em \PLambda}{}\xspace}

\def\to                 {\ensuremath{\rightarrow}\xspace}

\def\AT#1     {\ensuremath{A_{\mathrm{T}}^{#1}}\xspace}           

\def\C#1      {\ensuremath{\mathcal{C}_{#1}}\xspace}                       
\def\Cp#1     {\ensuremath{\mathcal{C}_{#1}^{'}}\xspace}                    
\def\Ceff#1   {\ensuremath{\mathcal{C}_{#1}^{\mathrm{(eff)}}}\xspace}        
\def\Cpeff#1  {\ensuremath{\mathcal{C}_{#1}^{'\mathrm{(eff)}}}\xspace}       
\def\Ope#1    {\ensuremath{\mathcal{O}_{#1}}\xspace}                       
\def\Opep#1   {\ensuremath{\mathcal{O}_{#1}^{'}}\xspace}                    

\newcommand{\tev}{\ifthenelse{\boolean{inbibliography}}{\ensuremath{~T\kern -0.05em eV}}{\ensuremath{\mathrm{\,Te\kern -0.1em V}}}\xspace}
\newcommand{\gev}{\ensuremath{\mathrm{\,Ge\kern -0.1em V}}\xspace}
\newcommand{\mev}{\ensuremath{\mathrm{\,Me\kern -0.1em V}}\xspace}
\newcommand{\kev}{\ensuremath{\mathrm{\,ke\kern -0.1em V}}\xspace}
\newcommand{\ev}{\ensuremath{\mathrm{\,e\kern -0.1em V}}\xspace}
\newcommand{\gevc}{\ensuremath{{\mathrm{\,Ge\kern -0.1em V\!/}c}}\xspace}
\newcommand{\mevc}{\ensuremath{{\mathrm{\,Me\kern -0.1em V\!/}c}}\xspace}
\newcommand{\gevcc}{\ensuremath{{\mathrm{\,Ge\kern -0.1em V\!/}c^2}}\xspace}
\newcommand{\gevgevcccc}{\ensuremath{{\mathrm{\,Ge\kern -0.1em V^2\!/}c^4}}\xspace}
\newcommand{\mevcc}{\ensuremath{{\mathrm{\,Me\kern -0.1em V\!/}c^2}}\xspace}

\def\gsim{{~\raise.15em\hbox{$>$}\kern-.85em
          \lower.35em\hbox{$\sim$}~}\xspace}
\def\lsim{{~\raise.15em\hbox{$<$}\kern-.85em
          \lower.35em\hbox{$\sim$}~}\xspace}

\def\pt         {\ensuremath{p_{\mathrm{ T}}}\xspace}

\def\degrees{\ensuremath{^{\circ}}\xspace}

\newcommand{\lum} {\ensuremath{\mathcal{L}}\xspace}

\def\tell1  {TELL1\xspace}
\def\ukl1   {UKL1\xspace}

\usepackage{tikz}
\usetikzlibrary{backgrounds,fit}
\usetikzlibrary{shapes.geometric, arrows}

\usepackage{mciteplus}

\DeclareGraphicsExtensions{.jpg,.pdf,.png}

\tikzstyle{startstop} = [rectangle, rounded corners, minimum width=3cm, minimum height=1cm, text centered, text width=3cm, draw=black, fill=black!5!white, font=\footnotesize]
\tikzstyle{process} = [rectangle, minimum width=3cm, minimum height=1cm, text centered, text width=3cm, draw=black, fill=black!5!white, font=\footnotesize]
\tikzstyle{decision} = [diamond, aspect=3, minimum width=0.8cm, minimum height=0.3cm, text centered, text width=2.5cm, draw=black, fill=white!90!yellow, font=\footnotesize, yshift=-0.2cm]

\tikzstyle{optional_decision} = [diamond, dashed, aspect=3, minimum width=0.8cm, minimum height=0.3cm, text centered, text width=2cm, draw=black, fill=white!90!yellow, font=\footnotesize, yshift=-0.2cm]
\tikzstyle{process_description} = [rectangle, minimum width=3cm, minimum height=1cm, text centered, text width=3cm, font=\footnotesize]
\tikzstyle{process_cpu} = [rectangle, minimum width=3cm, minimum height=1cm, text centered, text width=3cm, draw=black, fill=white!85!blue, font=\footnotesize]

\tikzstyle{arrow} = [thick,->,>=stealth]
\tikzstyle{dashed_arrow} = [thick,->,>=stealth,dashed]

\begin{document}\sloppy

\title{Allen: A high level trigger on GPUs for LHCb}


\author{R.~Aaij*\thanks{* Corresponding authors: raaij@cern.ch, dovombru@cern.ch, dcampora@cern.ch\vspace{3mm}}\and
  J.~Albrecht \and
  M.~Belous \and
  P.~Billoir \and
  T.~Boettcher \and
  A.~Brea~Rodr\'iguez \and
  D.~vom~Bruch* \and
  D.~H.~C\'ampora~P\'erez* \and
  A.~Casais~Vidal \and
  D.~C.~Craik \and
  P.~Fernandez~Declara \and
  L.~Funke \and
  V.~V.~Gligorov \and
  B.~Jashal \and
  N.~Kazeev \and
  D.~Mart\'inez~Santos \and
  F.~Pisani \and
  D.~Pliushchenko \and
  S.~Popov \and
  R.~Quagliani \and
  M.~Rangel \and
  F.~Reiss \and
  C.~S\'anchez~Mayordomo \and
  R.~Schwemmer \and
  M.~Sokoloff \and
  H.~Stevens \and
  A.~Ustyuzhanin \and
  X. Vilas\'is Cardona \and
  M.~Williams
}


\institute{R.~Aaij \and D.~H.~C\'ampora~P\'erez
\at Nikhef National Institute for Subatomic Physics, Amsterdam, Netherlands
\and
J.~Albrecht \and L.~Funke \and H.~Stevens
\at Fakult\"{a}t Physik, Technische Universit\"{a}t Dortmund, Dortmund, Germany
\and
M.~Belous \and N.~Kazeev \and S.~Popov \and A.~Ustyuzhanin
\at National Research University Higher School of Economics, Moscow, Russia
\and
M.~Belous \and N.~Kazeev \and  D.~Pliushchenko \and S.~Popov \and A.~Ustyuzhanin
\at Yandex School of Data Analysis, Moscow, Russia
\and
P.~Billoir \and D.~vom~Bruch \and V.~V.~Gligorov \and F.~Reiss \and R.~Quagliani
\at LPNHE, Sorbonne Universit\'{e}, Paris Diderot Sorbonne Paris Cit\'{e}, CNRS/IN2P3, Paris, France
\and
T.~Boettcher \and D.~C.~Craik \and M.~Williams
\at Massachusetts Institute of Technology, Cambridge, United States
\and
A.~Brea~Rodr\'iguez \and A.~Casais~Vidal \and D.~Mart\'inez~Santos
\at Instituto Galego de F\'{i}sica de Altas Enerx\'{i}as (IGFAE), Universidade de Santiago de Compostela, Santiago de Compostela, Spain
\and
D.~H.~C\'ampora~P\'erez
\at Faculty of Science and Engineering, Maastricht University, Maastricht, Netherlands
\and
P.~Fernandez~Declara \and F.~Pisani \and R.~Schwemmer 
\at European Organization for Nuclear Research (CERN), Geneva, Switzerland
\and
P.~Fernandez~Declara
\at Department of Computer Science and Engineering, University Carlos III of Madrid, Madrid, Spain
\and
B.~Jashal \and C.~S\'anchez~Mayordomo
\at Instituto de F\'isica Corpuscular, Centro Mixto Universidad de Valencia - CSIC, Valencia, Spain
\and
F.~Pisani
\at INFN Sezione di Bologna, Bologna, Italy
\and
F.~Pisani
\at Universit\`{a} di Bologna, Bologna, Italy
\and
D.~Pliushchenko
\at National Research University Higher School of Economics, Saint-Petersburg, Russia
\and 
S.~Popov \and A.~Ustyuzhanin
\at National University of Science and Technology MISIS, Moscow, Russia
\and
M.~Rangel
\at Instituto de F\'isica, Universidade Federal do Rio de Janeiro (UFRJ), Rio de Janeiro, Brazil
\and
M.~Sokoloff
\at University of Cincinnati, Cincinnati, OH, United States
\and
X. Vilas\'is Cardona
\at DS4DS, la Salle, Universitat Ramon Llull, Barcelona, Spain
}

\date{Received: 18 December 2019 / Accepted: 3 April 2020}

\maketitle

\begin{abstract}
We describe a fully GPU-based implementation of the first level trigger for the upgrade of the LHCb detector, due to start data taking in 2021. 
  We demonstrate that our implementation, named Allen, can process the 40~Tbit/s data rate of the upgraded LHCb detector and perform a wide variety of pattern recognition tasks. These include finding the trajectories of charged particles, finding proton-proton collision points, identifying particles as hadrons or muons, and finding the displaced decay vertices of long-lived particles. We further demonstrate that Allen can be implemented in around 500 scientific or consumer GPU cards, that it is not I/O bound, and can be
  operated at the full LHC collision rate of 30~MHz.
  Allen is the first complete high-throughput GPU trigger proposed for a HEP experiment.

\end{abstract}

\section{Introduction}
\label{intro}

The LHCb detector~\cite{LHCb-DP-2014-002} at CERN is currently being upgraded for Run 3 of the LHC. It is due to begin data taking in 2021 at an instantaneous luminosity of \lum $= 2\times 10^{33} \, \textrm{cm}^{-2} \, \textrm{s}^{-1}$, corresponding to an average of around 6 proton-proton ($pp$) collisions per LHC bunch crossing.
At this luminosity, the rates of beauty and charm hadrons, which are of interest for most LHCb analyses, reach the MHz level in the LHCb detector's geometrical acceptance~\cite{Fitzpatrick2014}. The majority of them decays into fully hadronic final states.
Thus, efficiently reducing the output data rate requires finding charged particle trajectories (tracking) at the  first level of the real-time reconstruction (trigger).

As most of LHCb's data comes from its tracking detectors, which are responsible for the majority of readout channels, the upgraded detector operates a triggerless readout, in which all subdetectors
are read out at the full bunch crossing rate of 30 MHz, or a maximum data rate of 40 Tbit/s. 
Event selection relies on two software stages. In the first stage, called HLT1, events are primarily selected using inclusive one- and two-track based algorithms, in some cases requiring the track to be identified as a muon. At this stage, the close to optimal alignment and calibration constants from the previous run are used. HLT1 allows for an efficient reduction of the event rate by a factor 30 to 60, depending on the desired working point. In the second stage, called HLT2, the detector is aligned and calibrated in near-real-time and the remaining events undergo offline-quality track reconstruction, full particle identification and track fitting. Because of the high signal rate, HLT2 does not only classify bunch crossings (events) as interesting or uninteresting. Rather in most cases HLT2 identifies a decay of interest and associates it to one of the reconstructed $pp$ collisions. Subsequently for most physics analyses HLT2 outputs a reduced event format one order of magnitude smaller than the raw data, consisting of only objects related to the decay of interest and the associated $pp$ collision, following the approach pioneered in Run 2~\cite{LHCb-DP-2016-001,Aaij:2019uij,LHCbCollaboration:2319756}. This approach relies on the near-real-time detector alignment and calibration to maintain the ultimate detector performance without the need for costly ``offline'' reprocessing of the data, and results in a total output data volume of 80~Gbit/s.

\begin{figure}
  \centering
  \includegraphics[width=0.3\textwidth]{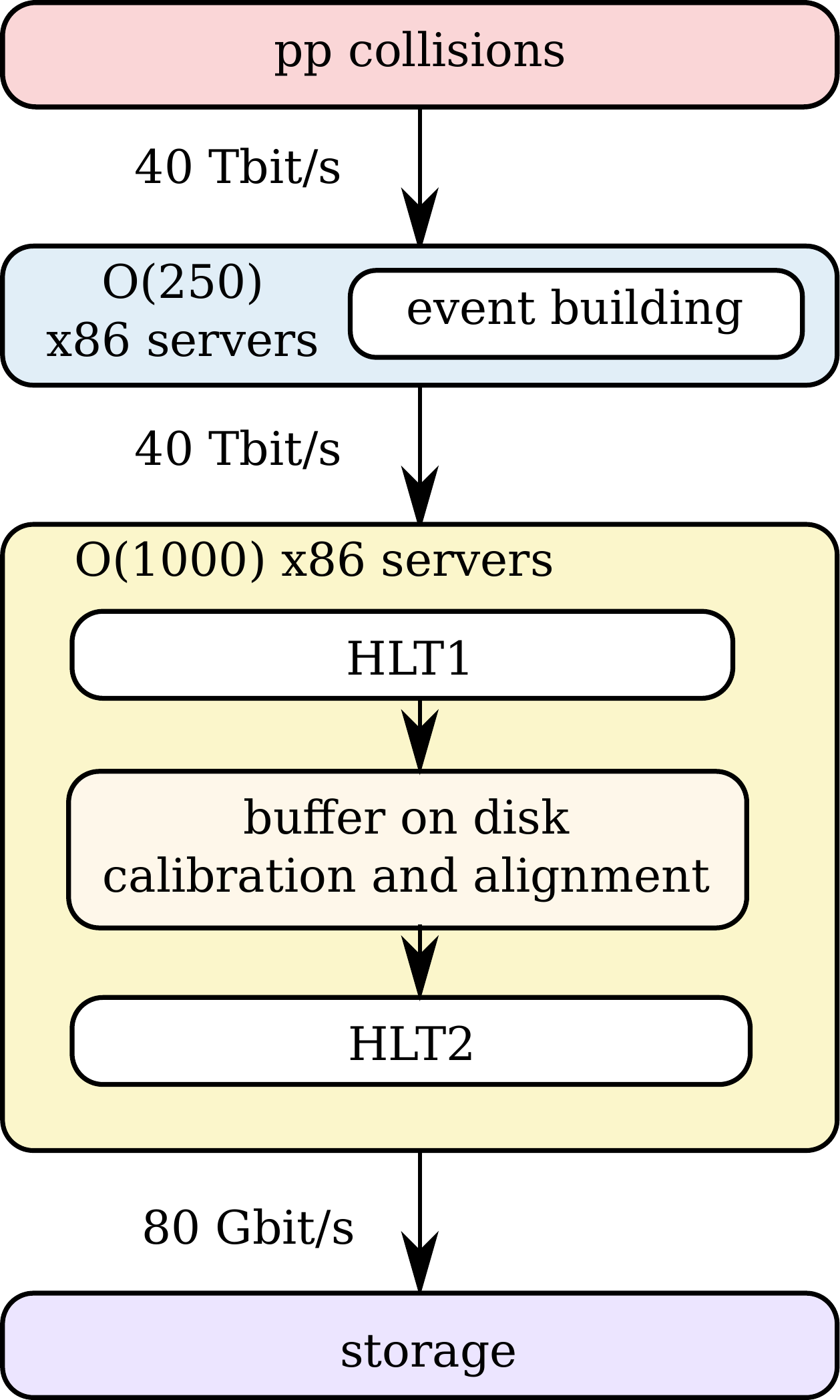}
  \caption{In the baseline proposal for the upgraded LHCb data acquisition system, x86 event building units receive data from the subdetectors and build events by sending and receiving event fragments over a 100G Infiniband (IB) network. The full data stream of built events is sent to x86 event filter servers to process both stages of the high level trigger.}
  \label{fig:daq}
\end{figure}

Performing full track reconstruction at 30~MHz and 40~Tbit/s poses a significant computing challenge. In the baseline proposal of the upgrade data acquisition system~\cite{Albrecht:2310579,LHCb-TDR-016}, data from the different LHCb subdetectors are received and combined to full events by about 250 event building x86 servers. Complete events are then sent to a separate ``event filter farm'' (EFF) of x86 servers, where both the HLT1 and HLT2 stages are
executed. Fig~\ref{fig:daq} shows this sequence of data processing units.

As track reconstruction is an inherently parallel problem, tracking algorithms can be designed to map well to the many-core architecture of graphics processing units (GPUs). 
Furthermore, GPUs map well onto LHCb's data acquisition architecture, because the event building servers which host the $\sim 500$  FPGA cards required to receive data from the detector at 30~MHz can also host two GPU cards each. Therefore, if the track reconstruction required for HLT1 could be processed with at most 500 GPUs, LHCb could execute HLT1 already inside the event building servers and reduce the data volume by a factor 30-60, significantly reducing the networking cost associated with sending data to the EFF.

In recent years, several particle physics experiments have studied the performance of track reconstruction on GPUs. So far only the ALICE experiment at CERN has employed GPUs in their trigger, where tracks from a single subdetector are reconstructed on the GPU, but data reduction occurs on x86 CPUs~\cite{Rohr_2012}.
All other R\&D efforts are intended for future experiments or upgrades. 
 In most proposals, data from a single sub-detector is analyzed on the GPU at a significantly lower data rate than 40~Tbit/s~\cite{Sen2015,Funke2014,VomBruch2017}. For some, the GPU coprocessor performs track reconstruction, but event selection or data reduction occur on x86 CPUs~\cite{Funke2014}. In other cases, event selection for a single physics signature runs on the GPU~\cite{Sen2015,VomBruch2017}. For Run 3, ALICE plans to perform track reconstruction of more than one subdetector and data compression on the GPU, at a data rate of 5~Tbit/s~\cite{Rohr2017}. 
 
In this paper, 
we show that for LHCb it is possible to execute a full trigger stage, including track reconstruction for several subdetectors and a variety of physics selections, at 40~Tbit/s on about 500 GPUs. We describe our implementation, named Allen 
after Frances E. Allen, following the LHCb convention of naming software projects after renowned scientists.

\section{Mapping the first trigger stage to graphics processing units}
\subsection{Characteristics of graphics processing units}
\begin{figure}
  \centering
  \includegraphics[width=0.4\textwidth]{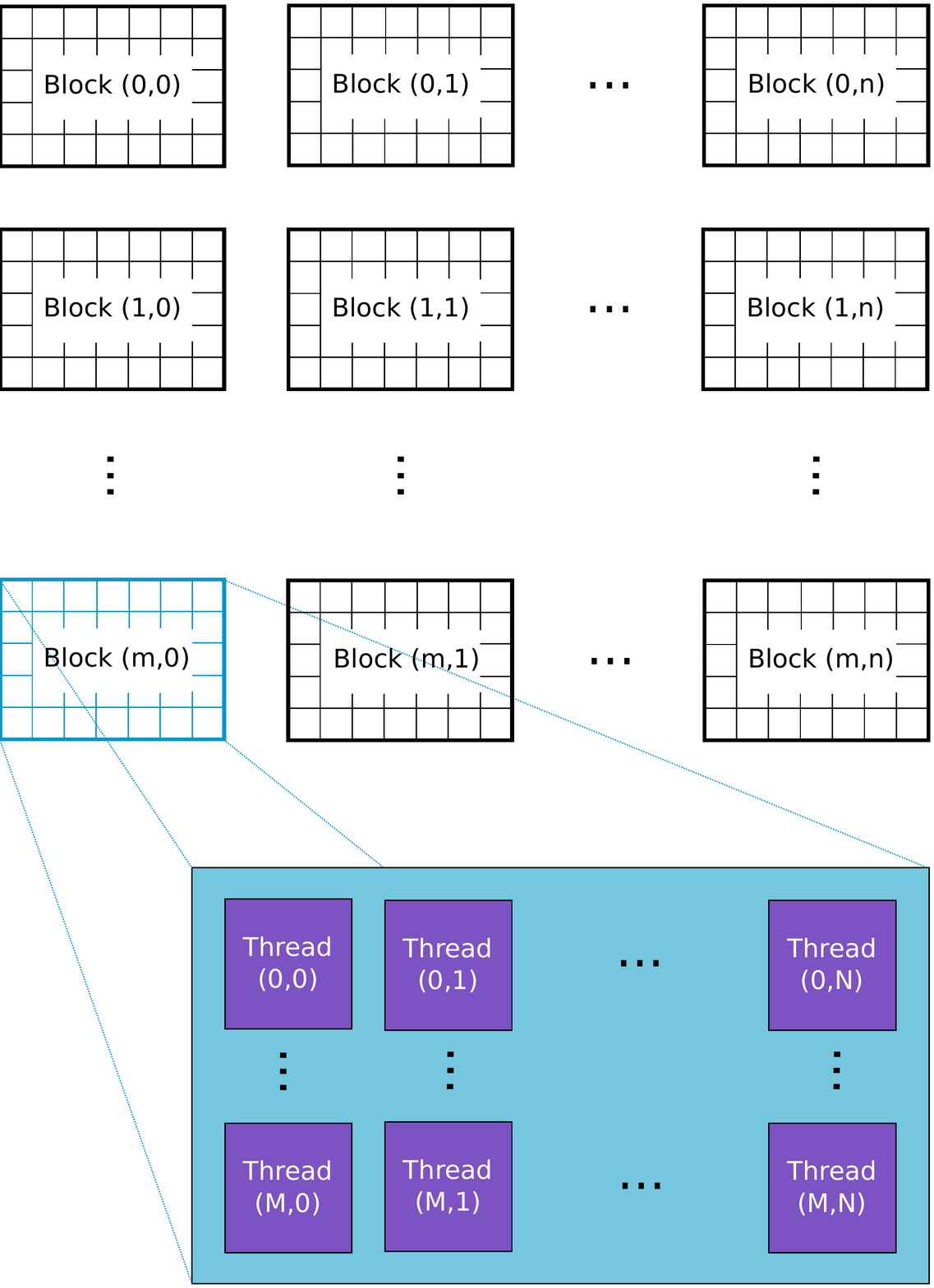}
  \caption{Threads are grouped into blocks, forming a grid that executes one kernel on the GPU.}
  \label{fig:gpu_threads_blocks}
\end{figure}

Developed for the graphics processing pipeline, GPUs excel at data parallel tasks under the SIMT paradigm~\cite{Lindholm2008}.
An algorithm executed on the GPU is called a kernel. Every kernel is launched with many threads on the GPU executing the same instruction on different parts of the data in parallel, independently from each other. These threads are grouped into blocks within a grid, as illustrated in Fig~\ref{fig:gpu_threads_blocks}. Threads within one block share a common memory and can by synchronized, while threads from different blocks cannot communicate. The threads are mapped onto the thousands of cores available on modern GPUs for processing.

Typically, a GPU is connected to its CPU host server via a PCIe connection, which sets a limit on the bandwidth between the GPU and the CPU: 16 lanes of PCIe 3.0 and PCIe 4.0 provide 128~Gbit/s and 256~Gbits/s, respectively. From these parameters we conclude that 500 GPUs are able to consume the 40~Tbit/s data rate of the upgraded LHCb detector. The total memory on a GPU is on the order of hundreds of Gbits nowadays. Consequently, 500 GPUs should also be able to process the full HLT1 sequence if enough data processing tasks fit into GPU memory at the same time and if the tasks can be sufficiently parallelized to fully unlock the TFLOPs theoretically available on the GPU.

\subsection{The Allen concept}

\begin{figure}
  \centering
  \includegraphics[width=0.3\textwidth]{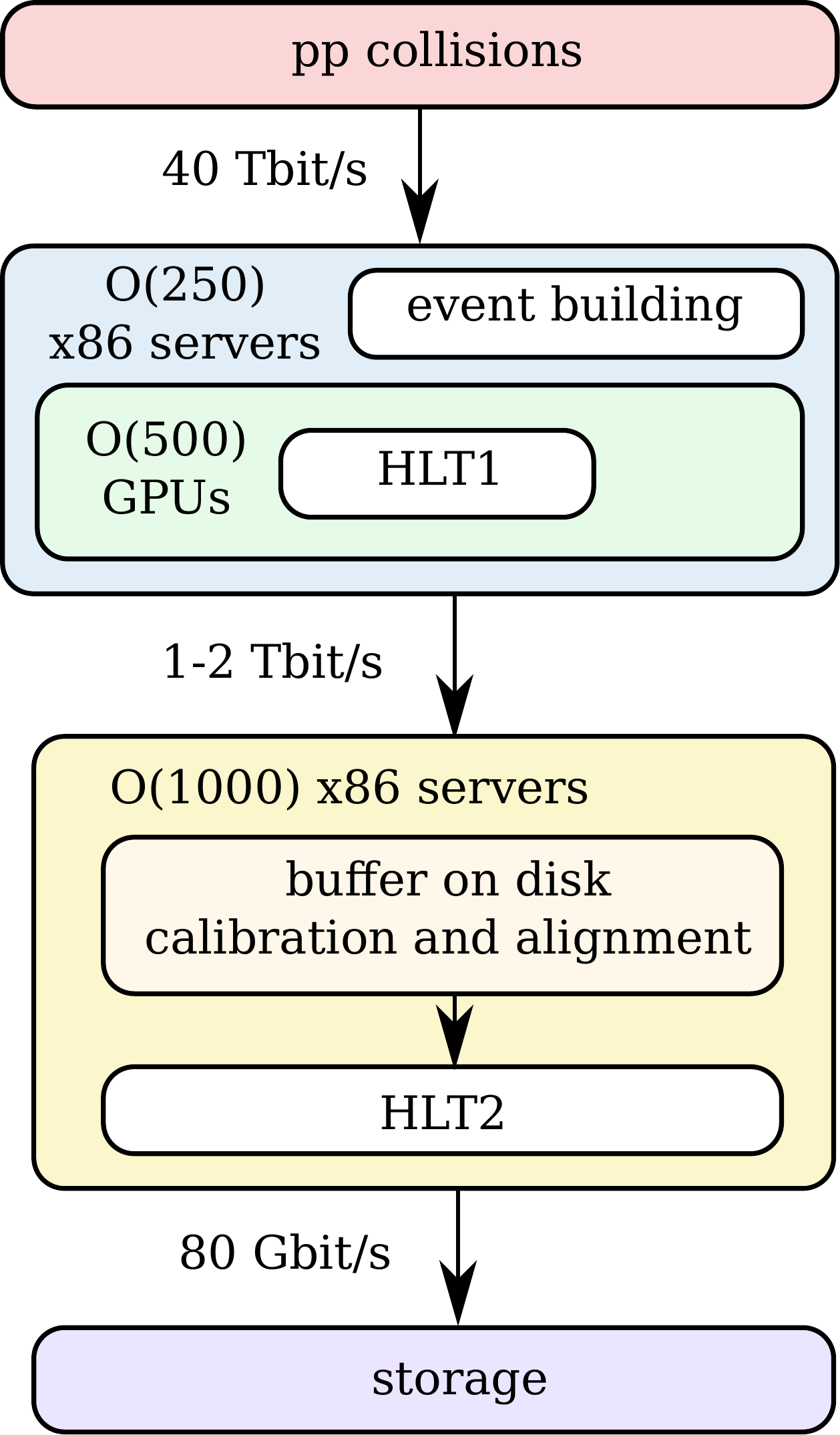}
  \caption{In the GPU-enhanced proposal for the upgraded LHCb data acquisition system x86 event building units receive data from the subdetectors and build events by sending and receiving event fragments over a 100G Infiniband (IB) network. The same x86 servers also host GPUs which process HLT1. Only events selected by HLT1 are sent to the x86 servers processing HLT2. The data rate between the two x86 server farms is therefore reduced by a factor 30 - 60.}
  \label{fig:daq_with_gpus}
\end{figure}

In our proposal, a farm of GPUs processes the full data stream, as shown in Fig~\ref{fig:daq_with_gpus}, which can be compared to the baseline x86-only architecture of Fig~\ref{fig:daq}. Every GPU receives complete events from an event building unit and handles several thousand events at once. Raw detector data is copied to the GPU, the full HLT1 sequence is processed on the GPU and only selection decisions and objects used for the selections, such as tracks and primary vertices, are copied back to the CPU.
This approach is motivated by the following considerations:

\begin{itemize}
\item LHCb raw events have an average size of 100 kB. When copying raw data to the GPU, the PCIe connection between the CPU and the GPU poses no limitation to the system, even when several thousand events are processed in parallel.

\item Since single events are rather small, several thousand events are required to make full use of the compute power of modern GPUs.

\item As the full algorithm sequence is processed on the GPU, no copies between the CPU and the GPU are required, apart from the raw input and selection output, and quantities needed to define the grid sizes of individual kernels.

\item Intra-GPU communication is not required because events are independent from one another and small enough in memory footprint to be processed on a single GPU.
  
\end{itemize}

The project is implemented in CUDA, Nvidia's API for programming its GPUs~\cite{cuda}. Allen\footnote{https://gitlab.cern.ch/lhcb/Allen Version 0.8 was used for the results in this publication.} includes a custom scheduler and GPU memory manager, which will be described in a companion publication.

\subsection{Main algorithms of the first trigger stage}
\begin{figure*}
  \centering
  \includegraphics[width=0.75\textwidth]{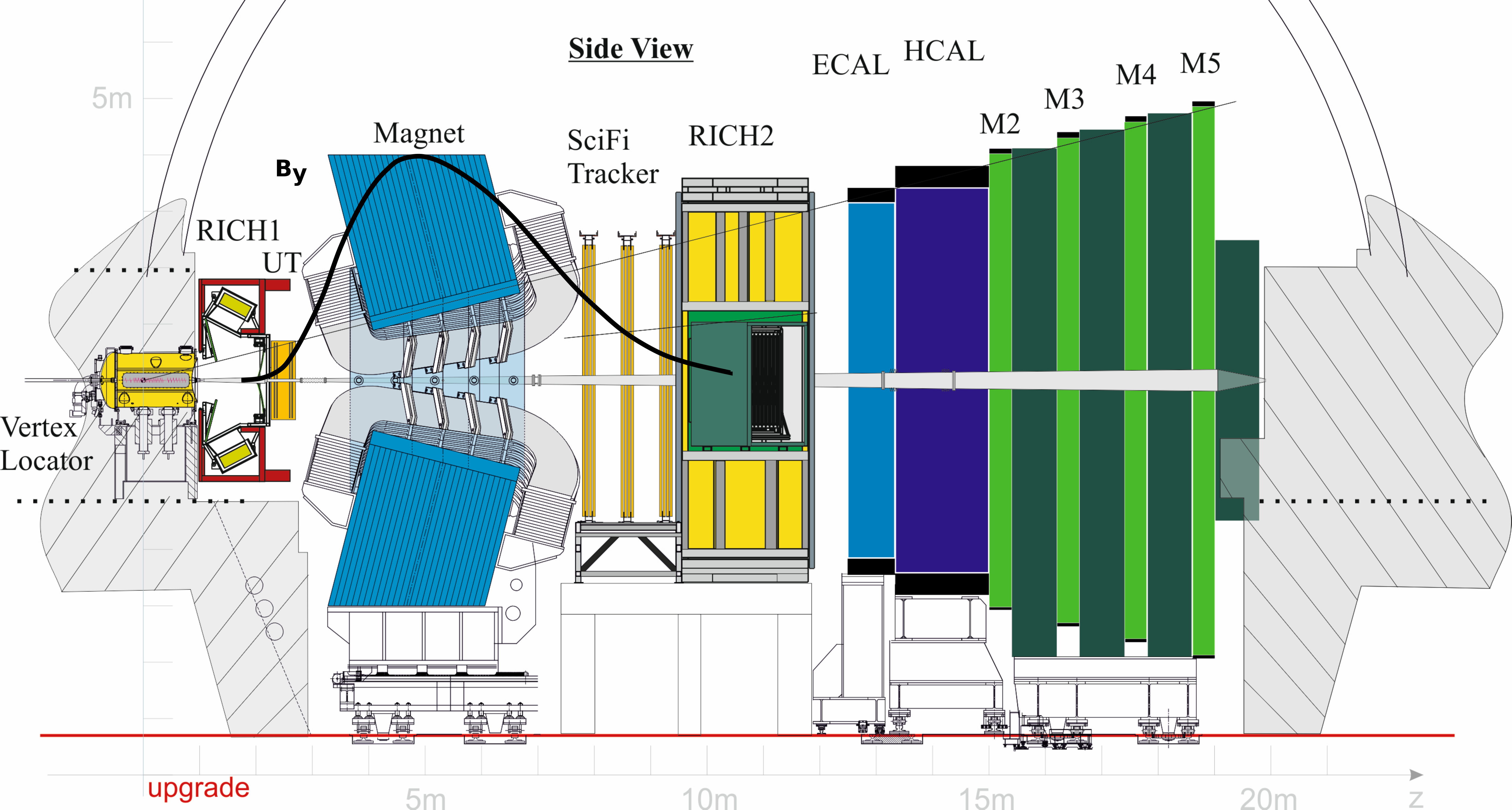}
  \caption{Upgraded LHCb detector. The y-component of the magnetic field $B_y$ is overlaid to visualize in which parts of the detector trajectories are bent. The maximum $B_y$ value is 1.05~T.}
  \label{fig:lhcb_detector}
\end{figure*}

A schematic of the upgraded LHCb forward spectrometer is shown in Fig~\ref{fig:lhcb_detector}. The information from the tracking detectors and the muon system is required for HLT1 decisions, as described in section~\ref{intro}. The tracking system consists of the vertex detector (Velo)~\cite{LHCb-TDR-013} and the upstream tracker (UT)~\cite{LHCb-TDR-015} before the magnet and tracking stations behind the magnet which are made of scintillating fibres (SciFi)~\cite{LHCb-TDR-015}. The measurements from the muon detector are used to perform muon identification. The LHCb coordinate system is such that \textbf{z} is along the beamline, \textbf{y} vertical and \textbf{x} horizonal. The dipole magnet bends charged particle trajectories along \textbf{x}.
Fig~\ref{fig:lhcb_detector} indicates the magnitude of the y-component of the magnetic field, which extends into the UT and SciFi regions. As a consequence, tracks in the Velo detector form straight lines, while those in the UT and SciFi detectors are slightly bent.

The following recurrent tasks are performed at various stages of the HLT1 sequence:

\begin{itemize}
\item Decoding the raw input into coordinates in the LHCb global coordinate system.
\item Clustering of measurements caused by the passage of the same particle into single coordinates (``hits''), depending on the detector type.
\item Finding combinations of hits originating from the same particle trajectory (pattern recognition).
\item Describing the track candidates from the pattern recognition step with a track model (track fitting).
\item Reconstructing primary and secondary vertices from the fitted tracks (vertex finding).
\end{itemize}

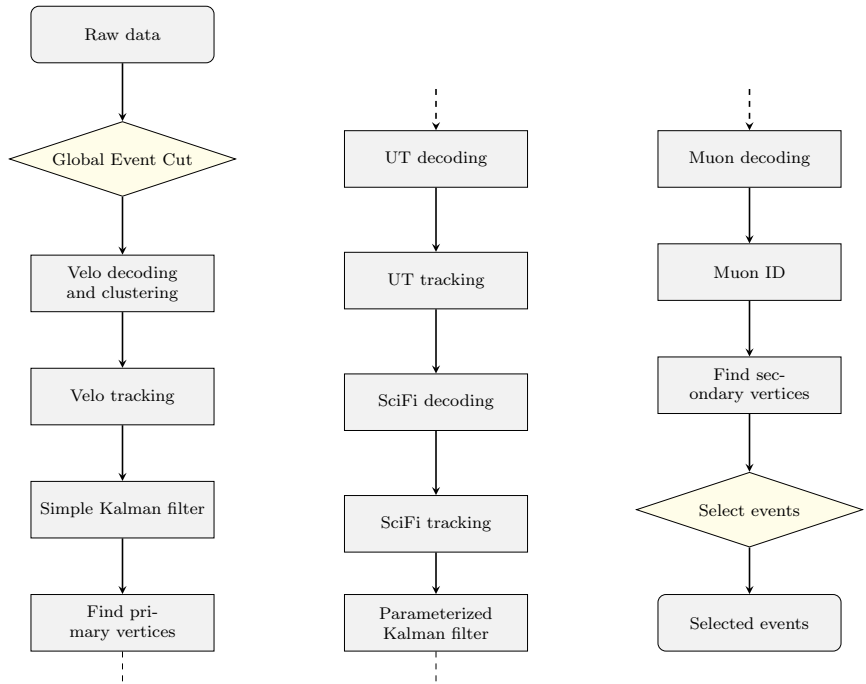
\begin{figure*}
  \centering
  \scalebox{0.75} {
    \begin{tikzpicture}[node distance=2cm]
      \node (initialize) [startstop] {Raw data};
      \node (gec) [decision, below of=initialize] {Global Event Cut};
      \node (velo_decoding) [process, below of=gec, yshift=-0.2cm] {Velo decoding and clustering};
      \node (velo_tracking) [process, below of=velo_decoding] {Velo tracking};
      \node (simple_kalman_filter) [process, below of=velo_tracking] {Simple Kalman filter};
      \node (find_primary_vertices) [process, below of=simple_kalman_filter] {Find primary vertices};
      \node (find_primary_vertices_arrow) [below of=find_primary_vertices, yshift=0.8cm] {};
      
      \node (ut_decoding) [process, right of=gec, xshift=3.5cm] {UT decoding};
      \node (ut_decoding_arrow) [above of=ut_decoding, yshift=-0.65cm] {};
      \node (ut_tracking) [process, below of=ut_decoding, yshift=-0.15cm] {UT tracking};
      \node (scifi_decoding) [process, below of=ut_tracking, yshift=-0.15cm] {SciFi decoding};
      \node (scifi_tracking) [process, below of=scifi_decoding, yshift=-0.15cm] {SciFi tracking};
      \node (kalman_filter) [process, right of=find_primary_vertices, xshift=3.5cm] {Parameterized Kalman filter};
      \node (kalman_filter_arrow) [below of=kalman_filter, yshift=0.8cm] {};
      
      \node (muon_decoding) [process, right of=ut_decoding, xshift=3.5cm] {Muon decoding};
      \node (muon_decoding_arrow) [above of=muon_decoding, yshift=-0.65cm] {};
      \node (muon_id) [process, below of=muon_decoding] {Muon ID};
      \node (find_secondary_vertices) [process, below of=muon_id] {Find secondary vertices};
      \node (select_events) [decision, below of=find_secondary_vertices] {Select events};
      \node (output) [startstop, right of=kalman_filter, xshift=3.5cm] {Selected events};
      
      \draw [arrow] (initialize) -- (gec);
      \draw [arrow] (gec) -- (velo_decoding);
      \draw [arrow] (velo_decoding) -- (velo_tracking);
      \draw [arrow] (velo_tracking) -- (simple_kalman_filter);
      \draw [arrow] (simple_kalman_filter) -- (find_primary_vertices);
      
      \draw [arrow] (ut_decoding) -- (ut_tracking);
      \draw [arrow] (ut_tracking) -- (scifi_decoding);
      \draw [arrow] (scifi_decoding) -- (scifi_tracking);
      \draw [arrow] (scifi_tracking) -- (kalman_filter);
      
      \draw [arrow] (muon_decoding) -- (muon_id);
      \draw [arrow] (muon_id) -- (find_secondary_vertices);
      \draw [arrow] (find_secondary_vertices) -- (select_events);
      \draw [arrow] (select_events) -- (output);
      
      \draw [dashed_arrow] (ut_decoding_arrow) -- (ut_decoding);
      \draw [dashed_arrow] (muon_decoding_arrow) -- (muon_decoding);
      \draw [dashed] (kalman_filter) -- (kalman_filter_arrow);
      \draw [dashed] (find_primary_vertices) -- (find_primary_vertices_arrow);
    \end{tikzpicture}
    }
\caption{Full HLT1 sequence implemented in CUDA to run on GPUs. Raw data is copied as input to the GPU, selected events are copied back to the host CPU as output. Rhombi represent algorithms reducing the event rate, while rectangles represent algorithms processing data.}
\label{fig:hlt1_sequence}
\end{figure*}
 
Fig~\ref{fig:hlt1_sequence} shows the full HLT1 sequence. In most cases, a single event is assigned to one block, while intra-event parallelism is mapped to the threads within one block. This ensures that communication is possible among threads processing the same event. Typically, the raw input is segmented by readout unit (for example a module of the vertex detector), so naturally the decoding can be parallelized among the readout units. During the pattern recognition step, many combinations of hits are tested and those are processed in parallel. The track fit is applied to every track and therefore parallelizable across tracks. Similarly, extrapolating tracks from one subdetector to the next is executed in parallel for all tracks. Finally, combinations of tracks are built when finding vertices and those can be treated in parallel.

Initially, events are preselected by a Global Event Cut (GEC) based on the size of the UT and SciFi raw data, removing the 10~\% busiest events.
This selection is not essential for the viability of the proposed GPU architecture. It is also performed in the baseline x86 processing~\cite{LHCb-TDR-016}, 
because very busy events have a less efficient detector reconstruction and their additional physics value to LHCb is not proportionate to the computing cost of reconstructing them. The subsequent elements of the HLT1 sequence are now described in turn.

\subsubsection{Velo detector}
The Velo detector consists of 26 planes of silicon pixel sensors placed around the interaction region. Its main purpose lies in reconstructing the $pp$ collisions (primary vertices or PVs) and in creating seed tracks to be further propagated through the other LHCb detectors.
The Velo track reconstruction is fully described in an earlier publication~\cite{CamporaPerez2019} and is recapped here for convenience.

The reconstruction begins by grouping measurements caused by the passage of a particle within each silicon plane into clusters, an example of a more general process known as connected component labeling. Allen uses a clustering algorithm employing bit masks, which searches for clusters locally in small regions. Every region can be treated independently, allowing for parallel processing. 

Straight-line tracks are reconstructed by first forming seeds of three hits from consecutive layers (``triplets''), and then extending these to the other layers in parallel. We exploit the fact that prompt particles produced in $pp$ collisions traverse the detector in lines of constant $\phi$ angle (within a cylindrical coordinate system where the cylinder axis coincides with the LHC beamline) and sort hits on every layer by $\phi$ for fast look-up when combining hits to tracks.

Velo tracks are fitted with a simple Kalman filter~\cite{Kalman1960} assuming that the x- and y-components are independent from one another and assigning a constant average transverse momentum of 400 MeV to all tracks for the noise contribution from multiple scattering. 

Finally, we search for PVs in a histogram of the point of closest approach of tracks to the beamline, where a cluster indicates a PV candidate. We refrain from a one-to-one mapping between a track and a vertex, which would introduce dependencies between the fitting of individual vertex candidates and would require sequential processing. Instead, every track is assigned to every vertex based on a weight, so that all candidates can be fitted in parallel. 

\subsubsection{UT detector}
Four layers of silicon strip detectors make up the UT detector, the strips of the two outer layers are aligned vertically, the two inner layers are tilted by +5\degrees and $-5\degrees$ around the z-axis respectively. Since more than 75\,\% of the hits consist of only one fired strip, no clustering is performed in this subdetector. The UT hits are decoded into regions based on their x-coordinate. Every region is then sorted by the y-coordinate. This allows for a fast look-up of hits around the position of an extrapolated Velo track. Velo tracks are extrapolated to the UT detector based on a minimum momentum cut-off of $3$~GeV, resulting in a maximal bending allowed between the Velo and UT detectors.There is no requirement on the transverse momentum. Subsequently, UT hits are assigned to Velo tracks and the track momentum is determined from the bending between the Velo and UT fitted straight-line track segments with a resolution of about 20\,\%. The UT decoding and tracking algorithms are described in more detail in Ref.~\cite{8756134}.

\subsubsection{SciFi detector}

The SciFi detector consists of three stations with four layers of scintillating fibres each, where the four layers of every station are in x-u-v-x configuration. The u- and v-layers are tilted by +5\degrees and $-5\degrees$, respectively, while the x-layers are vertical. The clustering of the SciFi hits and sorting along \textbf{x} is performed on the readout board; therefore, sorted clusters are obtained directly when decoding.

Tracks passing through both the Velo and UT detectors are extrapolated to the SciFi detector using a parameterization based on the track direction and the momentum estimate obtained after the UT tracking. This avoids loading the large magnetic field map into GPU memory. A search window defined by the UT track properties and a maximum number of allowed hits is determined for every UT track and every SciFi layer. 

The hit efficiency of the scintillating fibres is 98--99\%; therefore, several seeds are allowed per UT track, so that the track reconstruction efficiency is not limited by requiring hits from specific layers. Seeds are formed combining triplets of hits from within the search windows of one x-layer in each of the three SciFi stations. The curvature of tracks inside the SciFi region due to the residual magnetic field tails from the LHCb dipole is taken into account when selecting the best seeds. Only the seeds with the lowest $\chi^2$ relative to a parameterized description of the track within the SciFi volume are then extended by adding hits from the remaining x-layers, using the same track description. 
Since only the information of three hits is used for the $\chi^2$, its discriminating power is limited. Therefore, multiple track seeds are processed per UT track.

The magnetic field inside the SciFi detector can be expressed as $B_{y}(z) = B_{0}+B_{1}\cdot z$ and it is found that at first order $\frac{B_{1}}{B_{0}}$ is a constant. Using this parameterization, tracks are projected onto the remaining x and u/v-layers, and hits that deviate the least from the reference trajectory, within a track-dependent acceptance, are added. Only the U/V-layers  provide information on the track motion in the y-z plane. Thus, a parameterization accounting for the small curvature in the y-z plane 
is also taken into account in the track model, once all hits have been added.



Finally, a least means square fit is performed both in \vec{x} and \vec{y}. Every track is assigned a weight based on the normalized x-fit $\chi^2$, y-fit $\chi^2$, and the number of hits in the track. Only the best track is accepted per UT track, reducing fake tracks as much as possible.

\subsubsection{Muon detector}
The muon system~\cite{LHCb-TDR-014} consists of four multiwire proportional chambers interleaved with iron walls. Every station is divided into four regions with chambers of different granularity. Hits are read out with pads and strips, while strips from the same station can overlap to give a more accurate position measurement. During the decoding of muon measurements, such crossing strips are combined into a single hit. For muon identification, the ``isMuon'' algorithm described in Ref.~\cite{LHCb-DP-2013-001} is employed: tracks are extrapolated from the SciFi to the muon stations and muon hits are matched to a track within a region defined by the track properties. Depending on the track momentum, hits in different numbers of stations are required for a track to be tagged as a muon. 

\subsubsection{Kalman filter}
A Kalman filter is applied to all tracks to improve the impact parameter resolution, where the impact parameter (IP) is the distance between the point of closest approach of a track and a PV. The nominal LHCb Kalman filter uses a Runge-Kutta extrapolator to propagate track states between measurements and a detailed detector description to determine noise due to multiple scattering. In order to increase throughput and limit memory overhead, these costly calculations are replaced with parameterizations. Two versions of the parameterized Kalman filter are implemented in Allen: one which takes into account the whole detector and one which fits only the Velo track segment but using the estimated momentum from the full track passing through the Velo, UT and SciFi detectors. 
Since the impact parameter is mainly influenced by the measurements nearest to the interaction region, the Velo-only Kalman filter is used in the HLT1 sequence. This results in a significant computing speedup compared to applying the full Kalman filter.

\subsubsection{Selections}
Given the momentum, impact parameter and position information from the track fit as well as the muon identification, selections are applied on single tracks and two-track vertices similarly to the HLT1 selections used in Run2 of LHCb~\cite{Aaij:2652801,Likhomanenko:2015aba,BBDT}. Secondary vertices are fitted in parallel from combinations of two tracks each, providing a momentum and mass estimate for the hypothetical decaying particle, assigning the pion mass hypothesis to all tracks except for those identified as muons, for which the muon mass is assigned. The following five selection algorithms, which cover the majority of the LHCb physics programme and which are similar to lines accounting for about 95~\% of the HLT1 trigger rate in Run~2~\cite{Aaij:2652801}, are implemented in Allen:

\begin{itemize}
\item 1-Track: A single displaced track with $\pt > 1$~GeV.
\item 2-Track: A two-track vertex with significant displacement and $pt > 700$~MeV for both tracks.
\item High-$p_T$ muon: A single muon with $\pt > 10$~GeV for electroweak physics.
\item Displaced dimuon: A displaced dimuon vertex with $\pt > 500$~MeV for both tracks.
\item High-mass dimuon: A dimuon vertex with mass near or larger than the $J/\Psi$ mass with $\pt > 750$~MeV for both tracks.
\end{itemize}

\section{Results}
\begin{figure*}
  \centering
  \subfloat[]{\includegraphics[width=0.48\textwidth]{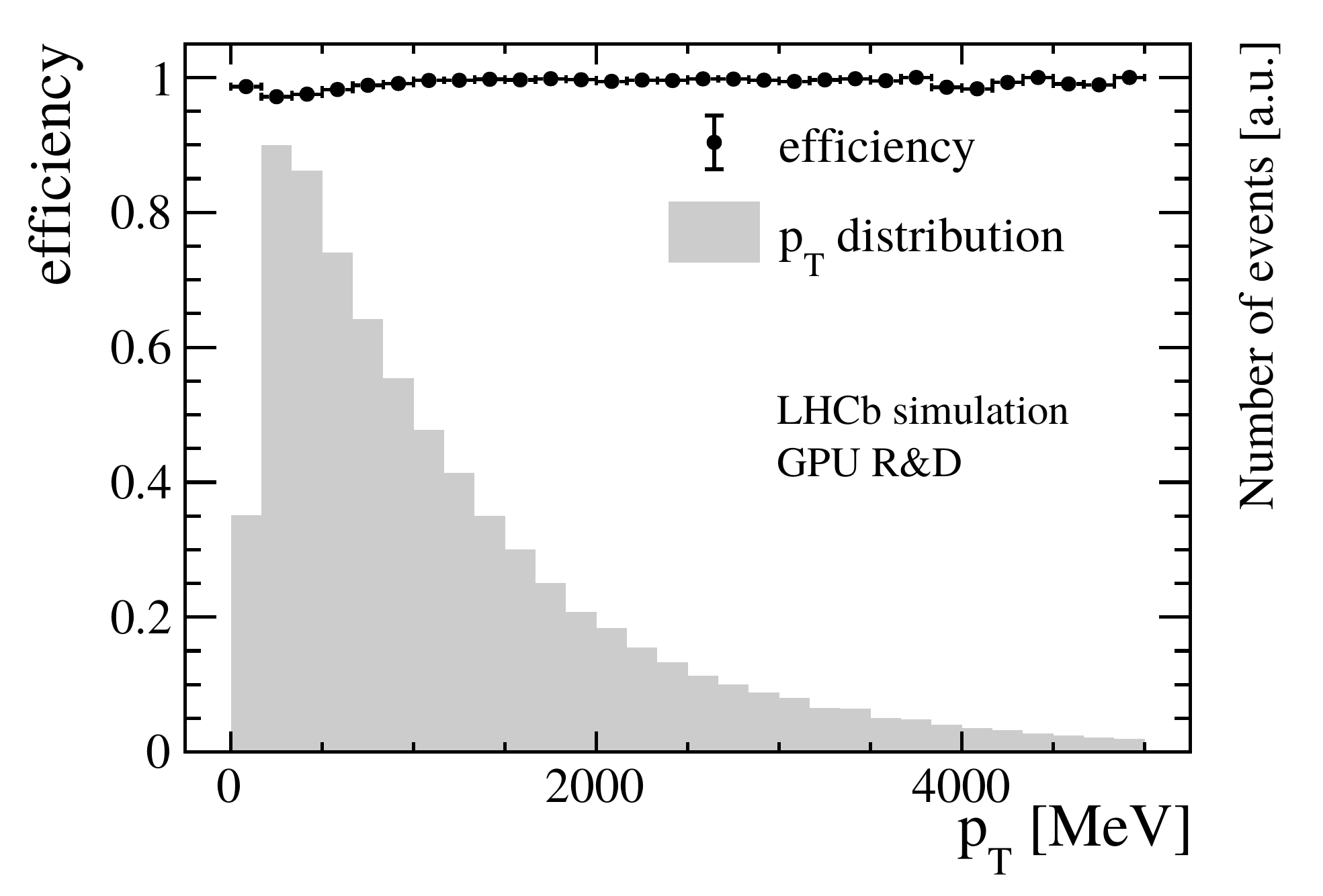}}
  \subfloat[]{\includegraphics[width=0.48\textwidth]{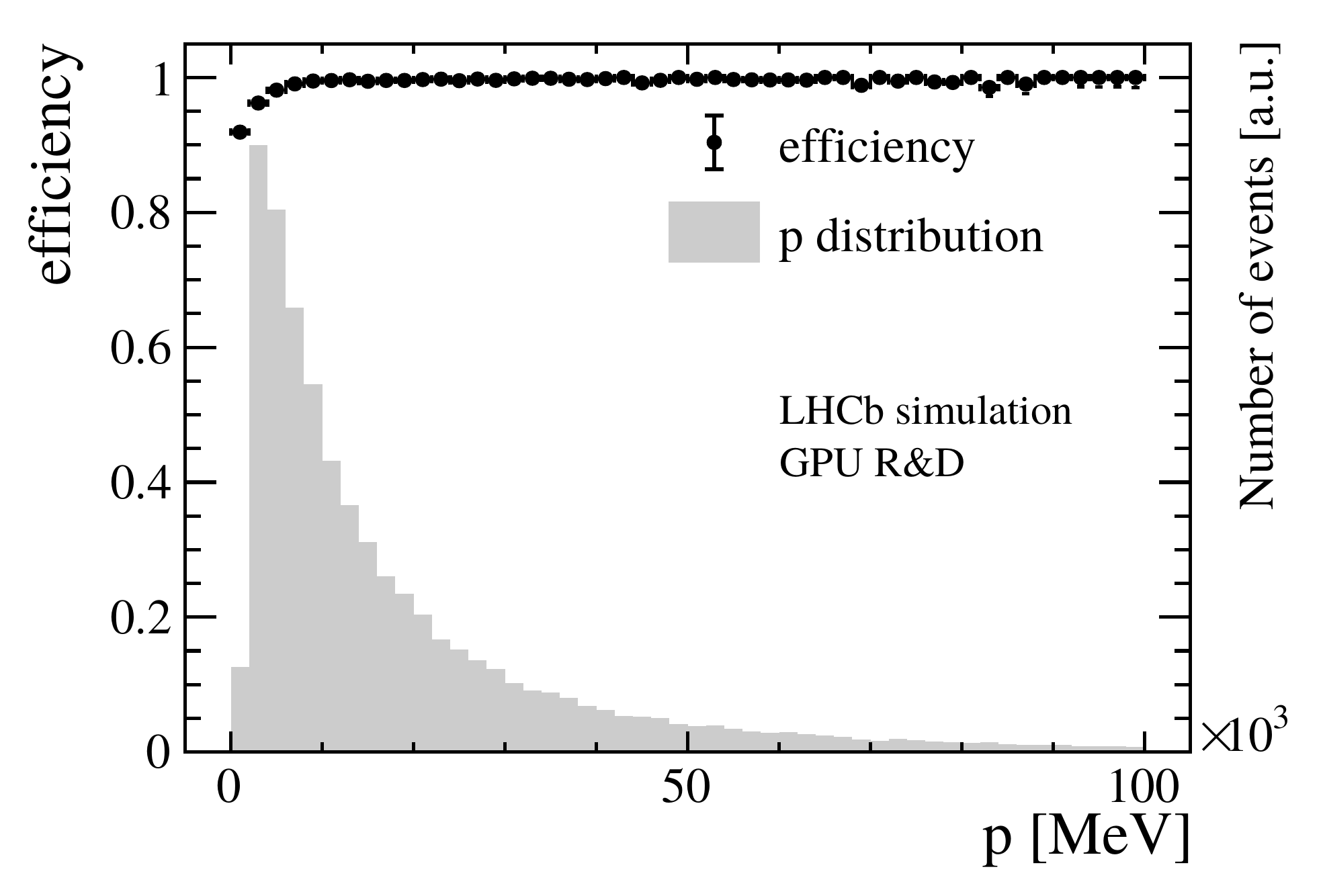}}\\
  \subfloat[]{\includegraphics[width=0.48\textwidth]{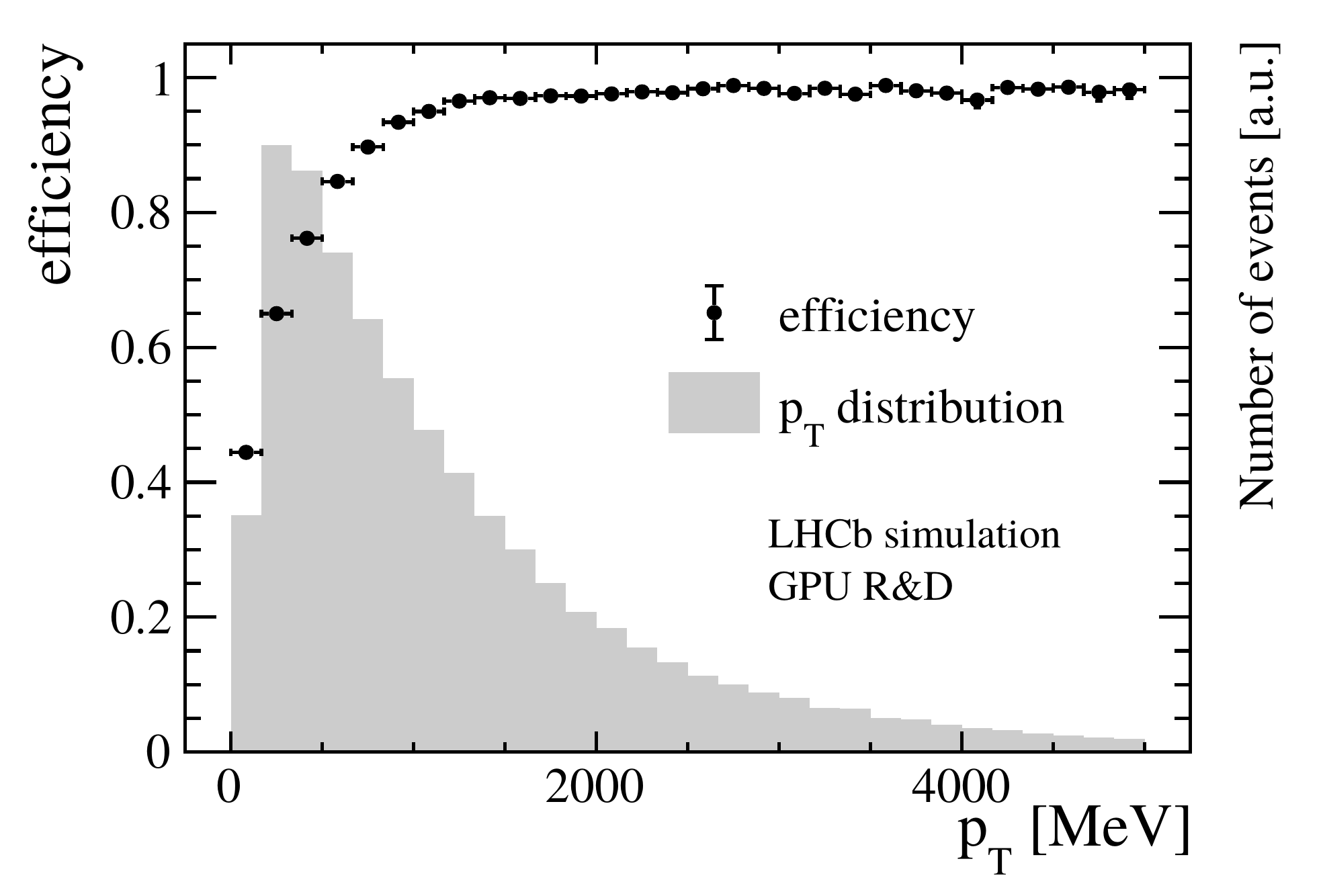}}
  \subfloat[]{\includegraphics[width=0.48\textwidth]{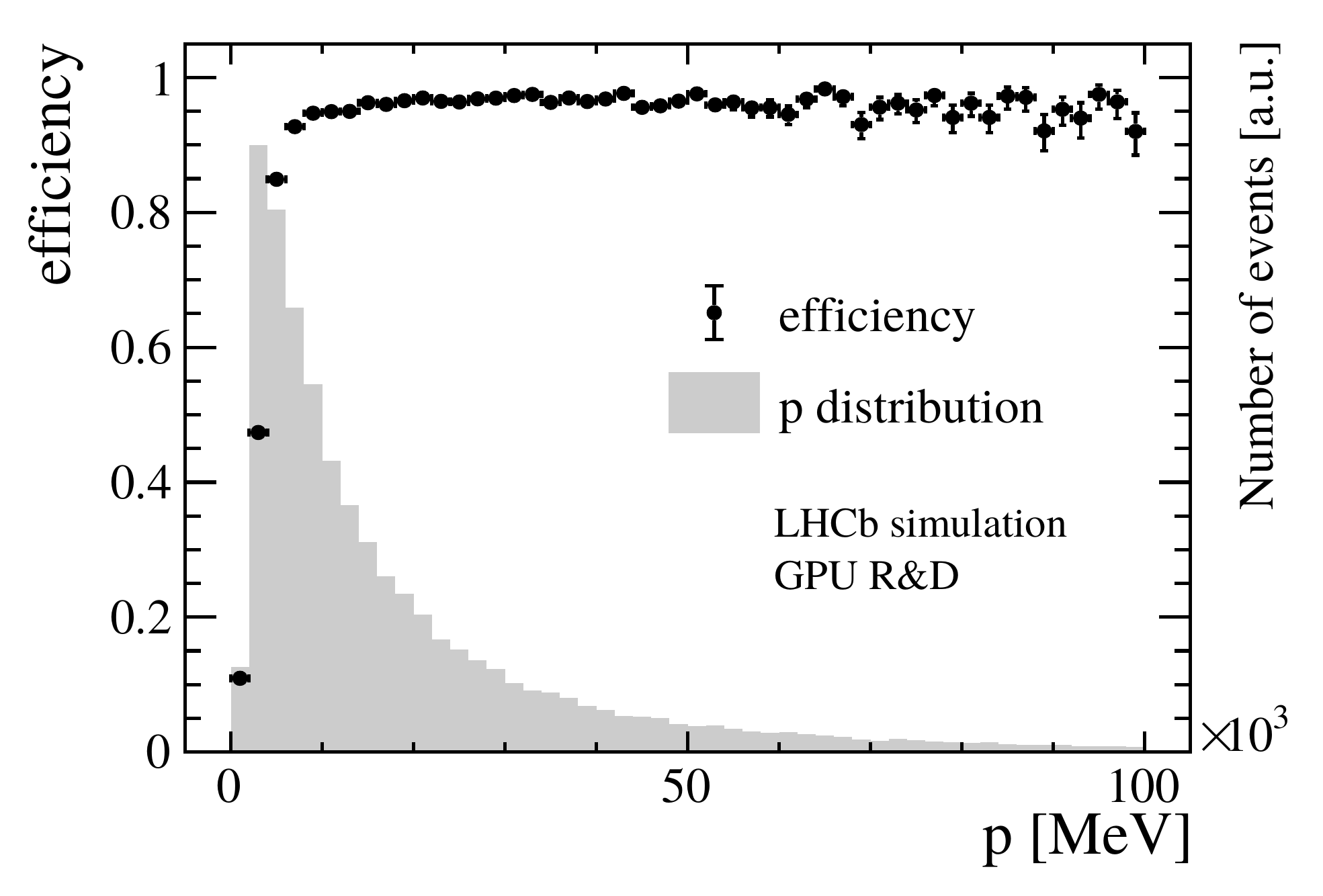}}\\
  \subfloat[]{\includegraphics[width=0.48\textwidth]{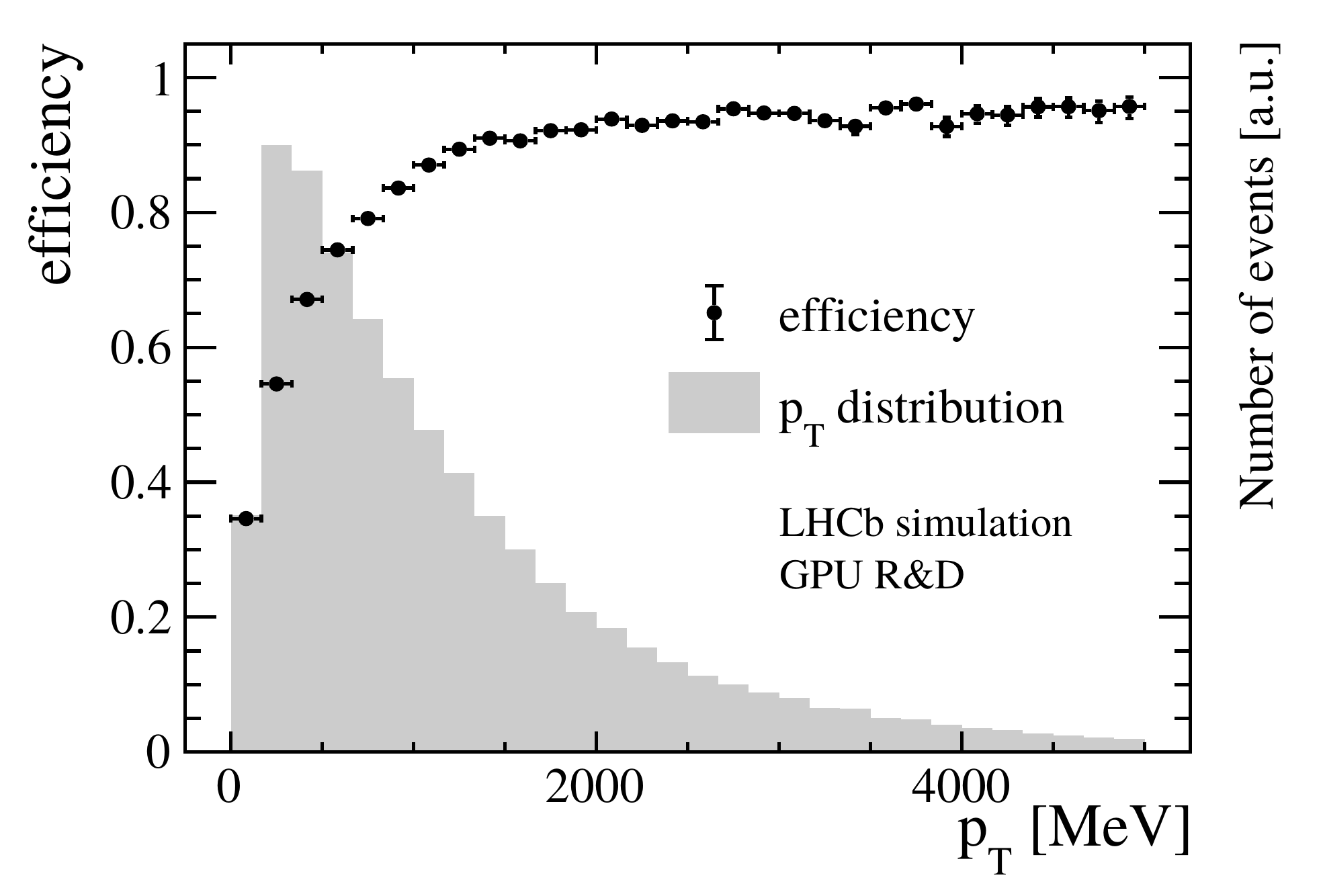}}
  \subfloat[]{ \includegraphics[width=0.48\textwidth]{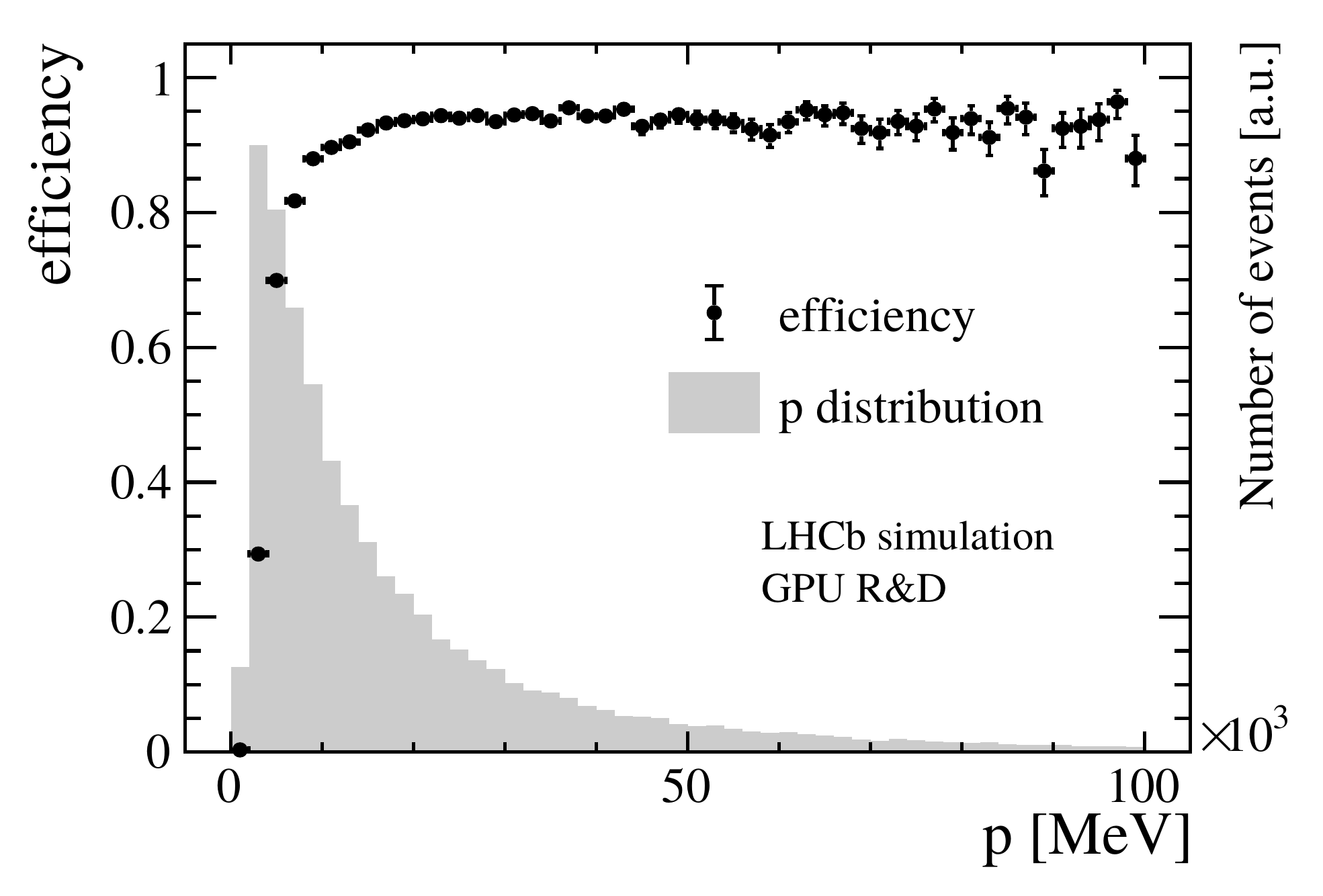}}
  \caption{Track reconstruction efficiency versus transverse momentum \pt (left) and momentum $p$ (right) of reconstructed tracks passing through the Velo (a, b), Velo and UT (c, d), Velo, UT and SciFi detectors (e, f) with respect to reconstructible non-electron tracks passing through the Velo, UT and SciFi detectors and produced from \B decays within the pseudorapidity coverage of the LHCb detector, $2 < \eta < 5$, for all signal samples combined. The \pt and $p$ distributions are overlaid as histograms.}
  \label{fig:efficiency_pt_eta}
\end{figure*}

The performance of Allen is studied both with respect to the computing throughput per GPU and the physics outcome in terms of track reconstruction efficiency and event selection efficiency for various representative LHCb analyses. 

\subsection{Physics performance}
For physics studies, simulated samples enhanced with decay channels of interest for the LHCb physics program are employed, namely a combination of 5000 events of each of the following decays: $\Bz\to\Kstarz\mumu$, $\Bz\to\Kstarz e^+e^-$, $\Bs\to\phi\phi$, $D^+_s \to \Kp\Km\pip$ and $Z\to\mu^+\mu^-$. Efficiencies of track and vertex reconstruction, muon identification and trigger selections, as well as the momentum resolution are determined directly within the Allen framework. 

\begin{figure}
  \centering
  \includegraphics[width=0.48\textwidth]{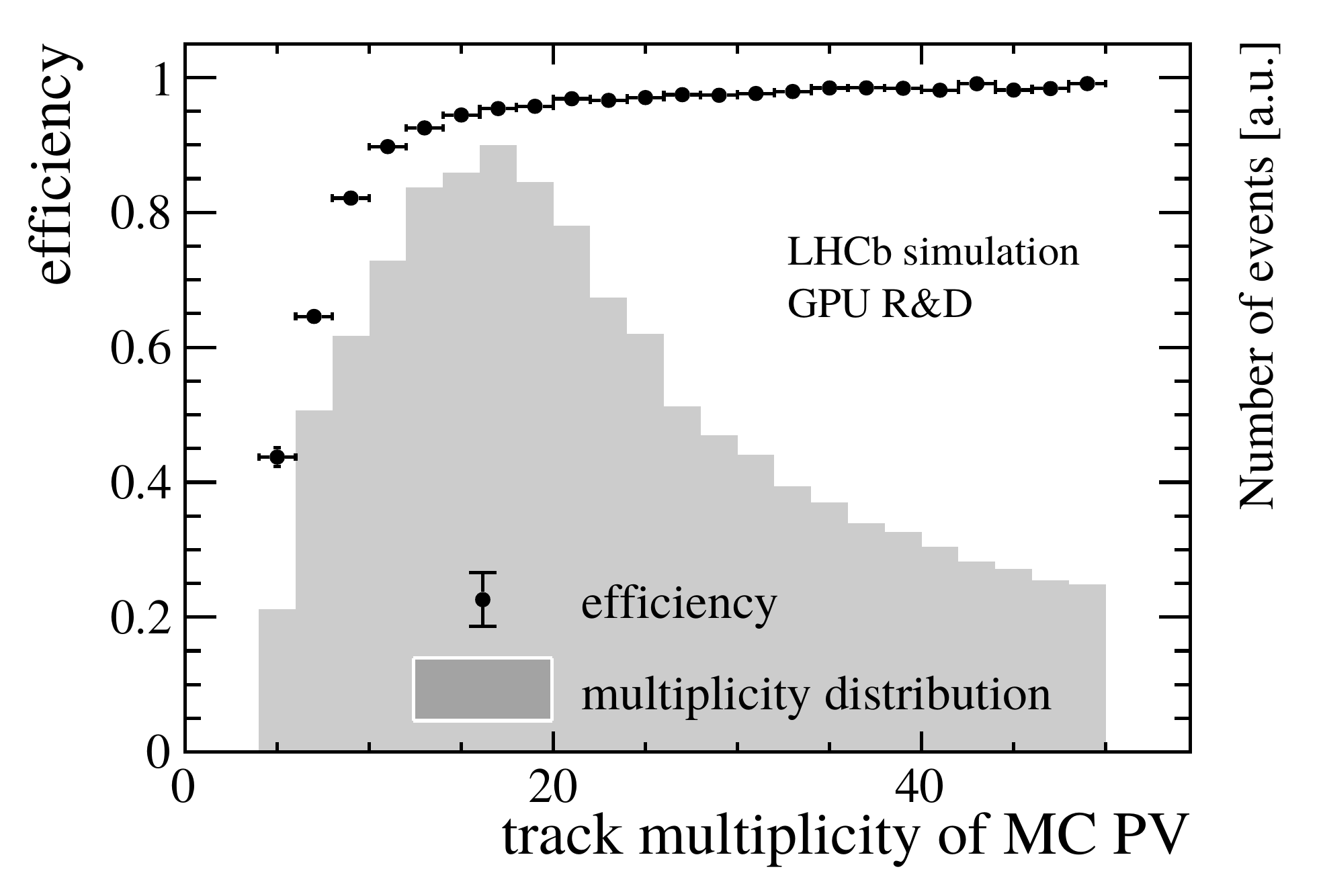}
  \caption{PV reconstruction efficiency versus track multiplicity of the MC PV for minimum bias events. The track multiplicity distribution is overlaid as a histogram.}
  \label{fig:pv_efficiency}
\end{figure}

\begin{figure}[ht]
  \centering
  \includegraphics[width=0.48\textwidth]{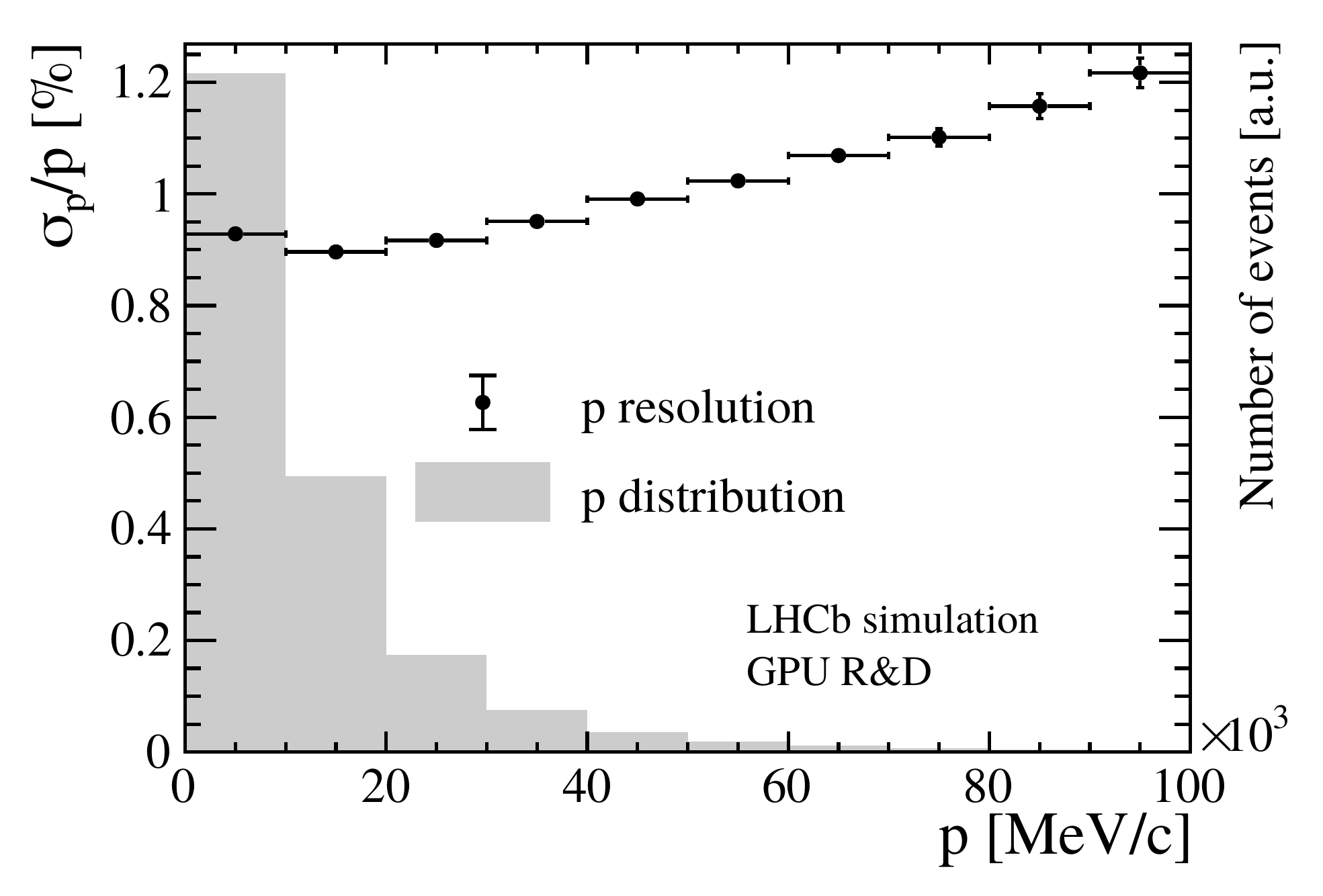}
  \caption{Relative momentum resolution of tracks passing through the Velo, UT and SciFi detectors versus momentum for all signal samples combined. Points represent the mean, error bars the width of a Gaussian distribution fitted to the resolution in every momentum slice. The momentum distribution is overlaid as a histogram.}
  \label{fig:scifi_momentum_resolution}
\end{figure}

\begin{figure}[ht]
  \centering
  \includegraphics[width=0.48\textwidth]{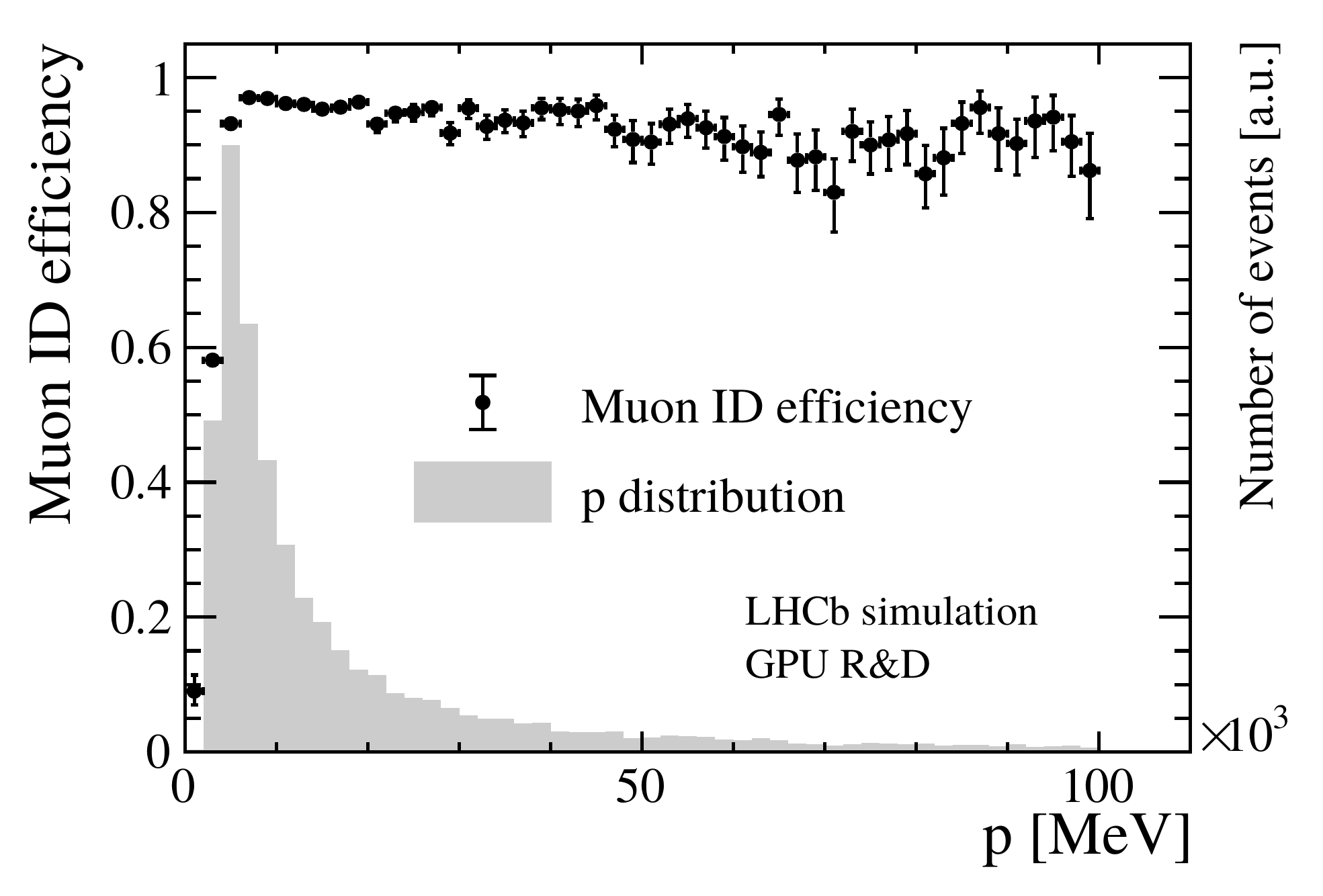}
  \caption{Muon identification efficiency versus momentum for tracks passing through the Velo, UT and SciFi detectors with respect to all reconstructible muons (explained in the text), for all signal samples combined. The momentum distribution is overlaid as a  histogram.}
  \label{fig:muon_id_efficiency}
\end{figure}

In LHCb, tracks are defined as correctly reconstructed if at least 70\,\% of the hits match those of the Monte Carlo (MC) particle associated to the track in simulation. Only MC particles resulting in the following minimum numbers of hits are considered as ``reconstructible tracks'': at least one hit in at least three different Velo modules and at least one hit in an x- and a u/v- layer in the UT detector and every station in the SciFi detector.
Fig~\ref{fig:efficiency_pt_eta} shows the track reconstruction efficiency of correctly reconstructed tracks in the Velo (top), Velo and UT (middle), Velo, UT and SciFi (bottom) detectors versus transverse momentum \pt and momentum $p$ with respect to reconstructible tracks originating from \B decays. 
A reconstructed PV is matched to a simulated PV if the distance is less than five times the uncertainty of the reconstructed PV along the z-axis. Fig~\ref{fig:pv_efficiency} shows the reconstruction efficiency of PVs versus the track multiplicity of the MC PV. As displayed in Fig~\ref{fig:scifi_momentum_resolution}, a relative momentum resolution better than 1\,\% is achieved which is sufficient for the selections of HLT1 and can be compared to a resolution of 0.5-1~\% obtained from offline-quality track reconstruction during Run~2. The muon identification efficiency is shown in Fig~\ref{fig:muon_id_efficiency}. It is determined with respect to "reconstructible muons", defined as reconstructed tracks which were matched to a muon MC particle.

\begin{figure}[ht]
  \centering
  \includegraphics[width=0.48\textwidth]{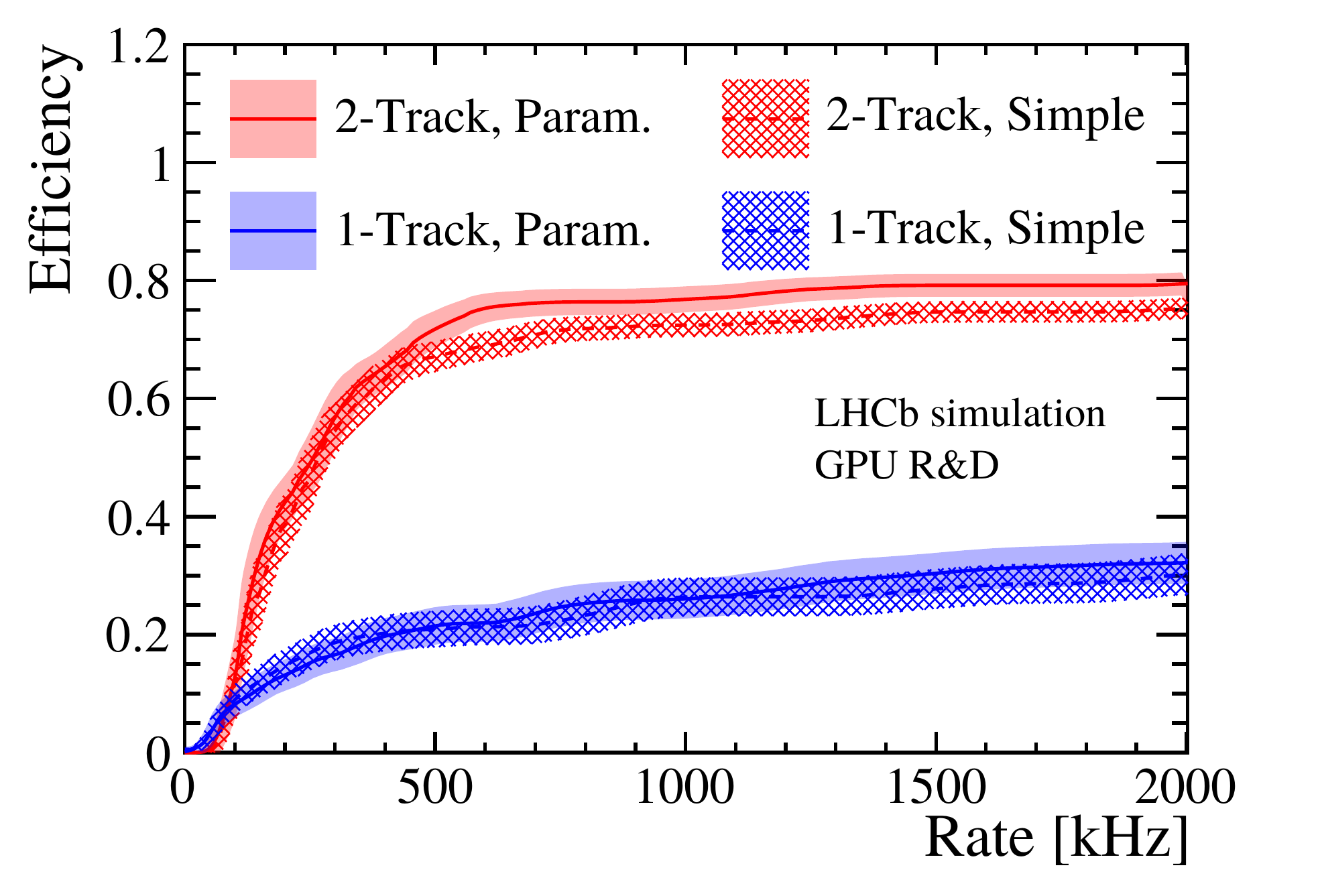}
  \caption{Efficiency of the 1-Track and 2-Track trigger lines when calculating the IP $\chi^2$ (see text for definition) from tracks fitted with the simple and parameterized Kalman filter, using the $\Bs\to\phi\phi$ sample. Varying the selection criteria of the IP $\chi^2$ results in rate and efficiency changes. The efficiency is calculated from subsets of the sample, the central value and error band correspond to the mean and standard deviation respectively.}
  \label{fig:efficiency_kalman_comparison}
\end{figure}

Finally, the trigger rates for the five selections  are shown in Table~\ref{tab:sigsels}. The total HLT1 output rate is about 1\,MHz, therefore, reducing the event rate by a factor 30. For this output rate, the selection efficiencies for various decay channels are given in Table~\ref{tab:sigeffs}. We quote the efficiency of the GEC, as well as for ``TIS'' events, with at least one passing trigger candidate not associated with a true signal decay product, and for ``TOS'' events, where the signal decay products must pass the trigger selection themselves.

Fig~\ref{fig:efficiency_kalman_comparison} illustrates the difference in efficiency and rate for the 1-Track and 2-Track trigger lines for the $\Bs\to\phi\phi$ sample between fitting tracks with the simple Kalman filter versus the parameterized Kalman filter, when varying the selection criteria of the IP $\chi^2$. The IP $\chi^2$ is defined as the difference between the $\chi^2$ of the PV reconstructed with and without the track under consideration and serves as estimate for the track displacement. Especially the efficiency of the 2-Track line improves when using the parameterized Kalman filter, since the momentum threshold for individual tracks is lower compared to the 1-Track line.

\begin{table}
  \begin{center}
  \begin{tabular}{l|c}
    \hline
    Trigger                    & Rate [kHz]\\
    \hline 
    1-Track                       & 215 $\pm$ 18 \\
    2-Track                  & 659 $\pm$ 31 \\
    High-$p_T$ muon            & 5 $\pm$ 3 \\ 
    Displaced dimuon           &  74 $\pm$ 10\\
    High-mass dimuon           & 134 $\pm$ 14 \\
    
    \hline 
    Total                      & 999 $\pm$ 38 \\
    \hline
   \end{tabular}
  \end{center}
  \caption{Rates of the five trigger selections implemented in Allen and the total HLT1 output rate, determined with minimum bias events.}
  \label{tab:sigsels}
\end{table}

\begin{table*}
  \begin{center}
  \begin{tabular}{l|cccc}
    \hline
    Signal                   & GEC & TIS -OR- TOS & TOS & $\mathrm{GEC}\times\mathrm{TOS}$\\
    \hline
    $\Bz\to\Kstarz\mumu$ & 88.9 $\pm$ 2.0 & 90.6 $\pm$ 2.0 & 88.8 $\pm$ 2.1 & 79.0 $\pm$ 2.6 \\
    $\Bz\to\Kstarz e^+e^-$ & 84.2 $\pm$ 2.7 & 69.1 $\pm$ 3.8 & 61.7 $\pm$ 4.0 & 52.0 $\pm$ 3.8 \\
    $\Bs\to\phi\phi$ & 83.2 $\pm$ 2.6 & 75.8 $\pm$ 3.2 & 68.5 $\pm$ 3.5 & 57.0 $\pm$ 3.4 \\
    $D^+_s \to \Kp\Km\pip$ & 82.5 $\pm$ 3.6 & 58.5 $\pm$ 5.1 & 42.6 $\pm$ 5.1 & 35.1 $\pm$ 4.5 \\
    $Z\to\mu^+\mu^-$ & 77.8 $\pm$ 1.2 & 99.5 $\pm$ 0.2 & 99.5 $\pm$ 0.2 & 77.4 $\pm$ 1.2 \\
    \hline
  \end{tabular}
  \end{center}
  \caption{Efficiencies of the total HLT1 selection. The TIS -OR- TOS and TOS efficiencies are calculated using events passing the GEC (definitions for TIS, TOS and GEC are in the text).  All efficiencies and their uncertainties are quoted in percentages and are determined from the different signal samples, with selections resulting in the rates given in Table~\ref{tab:sigsels}. Signal events are selected with the following criteria: b and c hadrons have a $\pt > 2$~GeV and a lifetime $\tau > 0.2$~ps. Children of b and c hadrons have $\pt > 200~$MeV. Children of Z bosons have $\pt > 20~$GeV.}
  \label{tab:sigeffs}
\end{table*}

\subsection{Computing performance}
For throughput studies, simulated samples of minimum bias events are used, representing the physics conditions expected for Run 3. The computing performance is compared among different Nvidia GPU cards. In all cases, Allen is compiled with gcc 8.2~\cite{gcc} and CUDA 10.1. The HLT1 sequence is run on a configurable number of concurrent threads. Each thread employs a GPU stream to asynchronously execute kernels and perform data transmission between CPU and GPU, 
such that memory transmissions do not impact throughput. The timer is started prior to processing a sequence in the first stream, and it is stopped after all streams have returned.

\begin{figure}[ht]
  \centering
  \includegraphics[width=0.48\textwidth]{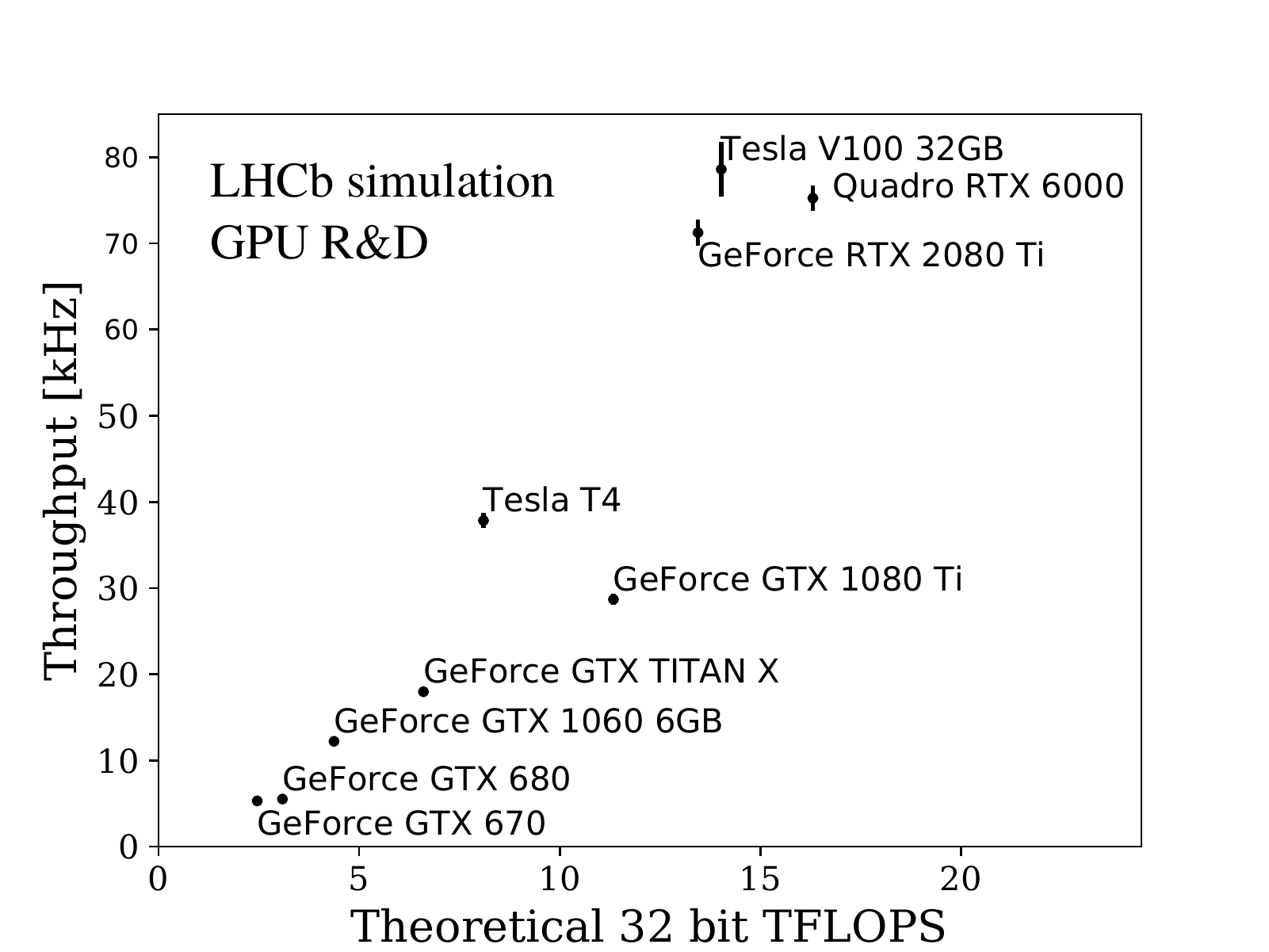}
  \caption{Allen throughput on various GPUs with respect to their reported peak 32-bit FLOPS performance. The mean and standard deviation of 10 measurements with different sets of 1000 events each are shown in the figure, with the measurement setup as described in the text.}
  \label{fig:hlt1perfvsflops}
\end{figure}

For most measurements, 12 thread-stream pairs with 1000 events each were processed 100 times, allocating 700~MB of GPU memory for every stream. Only in the case of the GTX 670, GTX 680 and the GTX 1060 6GB two thread-stream pairs were used instead. The measurement was performed 10 times with different sets of 1000 events each. The mean and standard deviation of the 10 measurements are shown for various Nvidia GPU cards as a function of their theoretical peak 32-bit FLOPS performance in Fig~~\ref{fig:hlt1perfvsflops}. The minimum rate per GPU necessary for processing the 30~MHz input rate with 500 GPUs is 60~kHz. Three cards surpass this threshold with a margin, namely the RTX 2080 Ti, the V100 and the Quadro RTX 6000, currently the best cards in the consumer, scientific and professional lines of Nvidia, respectively. Analyzing the performance as a function of theoretical peak 32-bit FLOPS performance reveals how the application scales to the hardware under study. The linear dependence visible in Fig~\ref{fig:hlt1perfvsflops} shows that the Allen code makes efficient use of the computing architecture and is likely to scale well to future generations of GPU processors.

\begin{figure}[ht]
  \centering
  \includegraphics[width=0.48\textwidth]{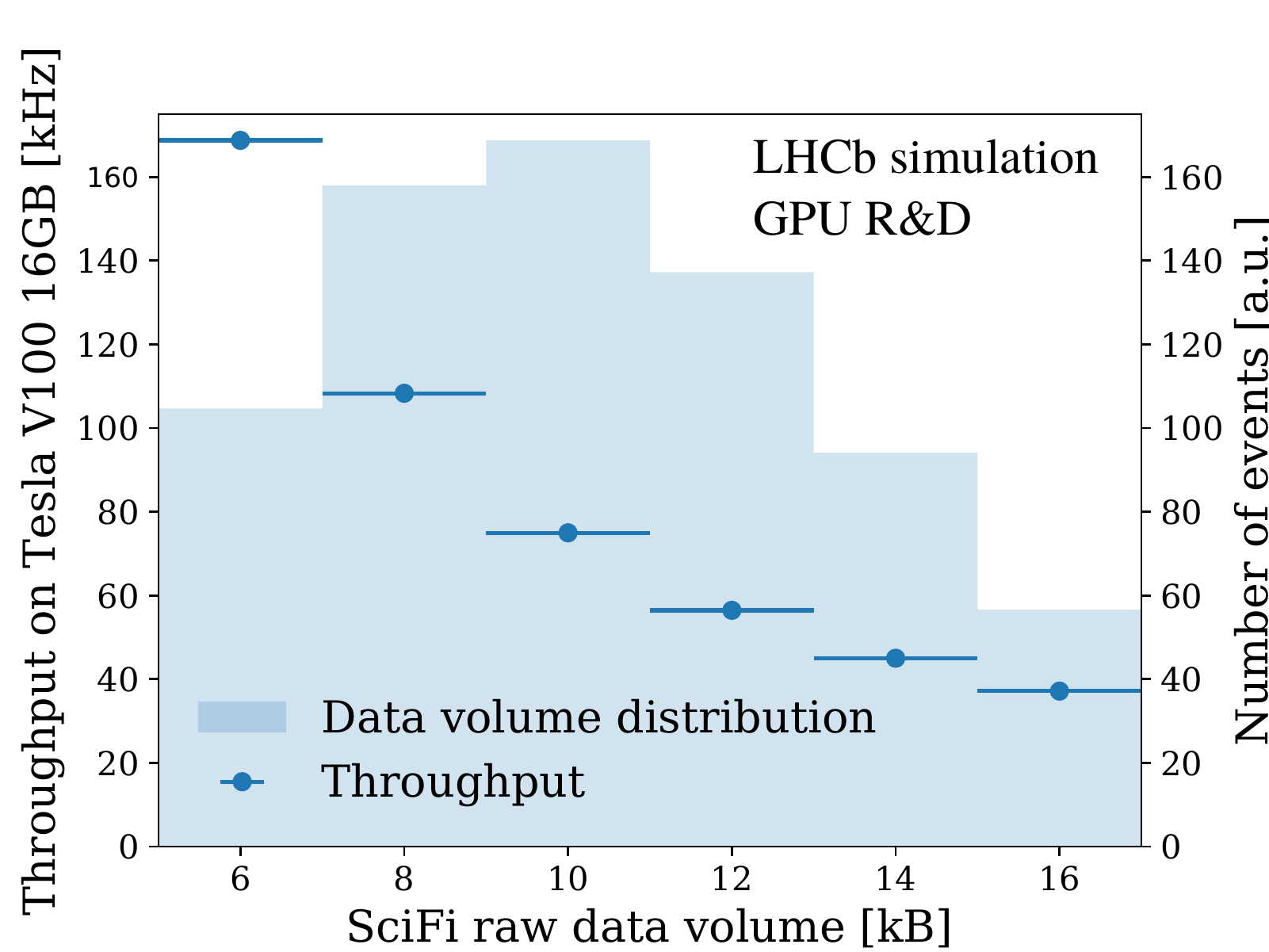}
  \caption{Throughput of the Allen sequence as a function of the SciFi raw data volume, which is proportional to the SciFi occupancy. The measurement setup is described in the text. For every data point, 1000 different events within the range of the SciFi raw data volume bin are processed. The GEC removing the 10\% busiest events was deactivated for these measurements.}
  \label{fig:throughputvsoccupancy}
\end{figure}

The throughput as a function of the occupancy in the SciFi detector is depicted in Fig~\ref{fig:throughputvsoccupancy}. The slower throughput decrease in the high occupancy region gives confidence that Allen can be adapted to real data taking conditions, where the detector occupancy might be higher than in simulation (as observed consistently during Runs 1 and 2). If the GEC removing the 10\% busiest events is deactivated and all events are processed, the Allen throughput drops by about 20\%.

\section{Conclusions}
We present Allen, an implementation of the first  trigger stage of LHCb for Run 3 entirely on GPUs. This is the first complete high-throughput GPU trigger proposed for a HEP experiment. Allen covers the majority of the LHCb physics programme, using an analogous reconstruction and selection sequence as in Run~2. The demonstrated event throughput shows that the full HLT1 sequence can run on about 500 of either one of the RTX 2080 Ti, V100 or Quadro RTX 6000 Nvidia GPU cards. Consequently, the GPUs can be hosted by the event building servers, significantly reducing the network cost associated with sending HLT1 output to the EFF. 
We show that the performance in terms of track and vertex reconstruction efficiencies, muon identification and momentum resolution are sufficient for efficient trigger selections for analyses representative of the LHCb physics programme. 

\section*{Acknowledgements}
We would like to thank N.~Neufeld and T.~Colombo for many fruitful discussions.  We  also  thank  the  LHCb  RTA team for supporting this publication and reviewing this work. We thank the technical and administrative staff at the LHCb institutes. We acknowledge support from CERN and from the national agencies: CAPES, CNPq, FAPERJ and FINEP (Brazil); MOST and NSFC (China); CNRS/IN2P3 (France); BMBF, DFG and MPG (Germany); INFN (Italy); NWO (Netherlands); MNiSW and NCN (Poland); MEN/IFA (Romania); MSHE (Russia); MinECo (Spain); SNSF and SER (Switzerland); NASU (Ukraine); STFC (United Kingdom); DOE NP and NSF (USA). We acknowledge the computing resources that are provided by CERN, IN2P3 (France), KIT and DESY (Germany), INFN (Italy), SURF (Netherlands), PIC (Spain), GridPP (United Kingdom), RRCKI and Yandex LLC (Russia), CSCS (Switzerland), IFIN-HH (Romania), CBPF (Brazil), PL-GRID (Poland) and OSC (USA). We are indebted to the communities behind the multiple open-source software packages on which we depend. Individual groups or members have received support from AvH Foundation (Germany); EPLANET, Marie Sk\l{}odowska-Curie Actions and ERC (European Union); ANR, Labex P2IO and OCEVU, and R\'{e}gion Auvergne-Rh\^{o}ne-Alpes (France); Key Research Program of Frontier Sciences of CAS, CAS PIFI, and the Thousand Talents Program (China); RFBR, RSF and Yandex LLC (Russia); GVA, XuntaGal and GENCAT (Spain); the Royal Society and the Leverhulme Trust (United Kingdom). J. Albrecht acknowledges support of the European Research Council Starting grant PRECISION 714536.
D. vom Bruch, V. V. Gligorov, F. Reiss and R.~Quagliani acknowledge support of the European Research Council Consolidator grant RECEPT 724777. H. Stevens, L. Funke acknowledge support of the Collaborative Research Center SFB 876. 
T.~Boettcher and M.~Williams are supported by US NSF grant PHY-1912836. 
D.~Craik is supported by US NSF grants OAC-1836650 and PHY-1904160. 
A.~Ustyuzhanin is supported by the Russian Science Foundation grant agreement n$^{\circ}$ 19-71-30020. D.~Mart\'inez~Santos and A.~~Brea~Rodr\'iguez acknowledge support from the European Research Council Starting grant BSMFLEET 639068.\vspace{1mm}
\section*{Conflict of Interest}
The authors declare that they have no conflict of interest.


%
%

\bibliographystyle{LHCb}
\bibliography{LHCb-CONF,LHCb-DP,LHCb-PAPER,LHCb-TDR,main}   


\end{document}